\documentclass[10pt]{article}

\usepackage{graphicx}
\usepackage{boxedminipage}
\usepackage{amsmath}
\usepackage{amssymb}
\usepackage{amscd}  
\usepackage{amsfonts}
\usepackage{amsthm}

\theoremstyle{plain}
\newtheorem{theorem}{Theorem}
\newtheorem{lemma}[theorem]{Lemma}

\theoremstyle{definition}
\newtheorem{definition}{Definiton}[section]

\theoremstyle{remark}

\newcommand{\mycaption}[1]{\caption{{\em #1}}}
\newcommand{\be}{\begin{eqnarray}}
\newcommand{\ee}{\end{eqnarray}}

\renewcommand{\d}{\operatorname{d}}   

\let\Par\S
\renewcommand{\S}{\operatorname{S}}   

\newcommand{\A}{{\mathcal A}}

\newcommand{\Aut}{\operatorname{Aut}}
\newcommand{\B}{{\mathcal B}}
\newcommand{\C}{{\mathcal C}}
\newcommand{\cf}{{\em cf.}~}

\newcommand{\eg}{{\em e.g.}~}
\newcommand{\etc}{{\em etc}}
\newcommand{\eq}{{\,  := \, }}
\newcommand{\End}{\operatorname{End}}

\newcommand{\F}{\mathcal{F}}

\newcommand{\G}{{\mathcal G}}

\newcommand{\Gset}{\mathbb{G}}

\newcommand{\half}{\frac{1}{2}}
\newcommand{\hol}{\operatorname{hol}}
\newcommand{\Hom}{\operatorname{Hom}}

\newcommand{\ie}{{\em i.e.}~}

\newcommand{\im}{\operatorname{im}}
\newcommand{\Inn}{\operatorname{Int}}

\newcommand{\Lie}{\operatorname{Lie}}

\newcommand{\OMEGA}{\underline{\Omega}}
\newcommand{\Out}{\operatorname{Out}}

\newcommand{\qq}{\mathrm{q}}
\newcommand{\QQ}{\bar{\mathrm{q}}}
\newcommand{\qqs}{\hat{\mathrm{q}}}
\newcommand{\qqq}{\mathrm{s}}
\newcommand{\Qn}{\mathrm{Q}}
\newcommand{\qqbarH}{{\bar{\mathrm{q}}^{\scriptscriptstyle{\mathrm{H}}}}}
\newcommand{\qqH}{{{\mathrm{q}}^{\scriptscriptstyle{\mathrm{H}}}}}\newcommand{\qqV}{{{\mathrm{q}}^{\scriptscriptstyle{\mathrm{V}}}}}

\newcommand{\Rset}{\mathbb{R}}

\newcommand{\tr}{\operatorname{tr}}
\newcommand{\Tors}{\operatorname{Tors}}
\newcommand{\Tr}{\operatorname{Tr}}

\newcommand{\unit}{\mathbf{1}}
\newcommand{\veva}{\tilde{a}}

\newcommand{\Z}{\operatorname{Z}}

\begin{document}

\begin{titlepage}
\begin{flushleft}
\hfill Imperial/TP/05/JK/01 \\
\hfill {\tt hep-th/0510069} \\
\hfill \\
\hfill Revised March 20, 2006 
\end{flushleft}

\vspace*{10mm}

\begin{center}
{\bf \Large Topological Quantum Field Theory on \\
\vspace{2mm} 
Non-Abelian Gerbes } \\
\vspace*{8mm} 

{Jussi Kalkkinen} \\

\vspace*{3mm}

{\em The Blackett Laboratory, Imperial College} \\
{\em Prince Consort Road, London SW7 2BZ, U.K.} \\

\vspace*{6mm}

\end{center}

\begin{abstract} 
The infinitesimal symmetries of a fully decomposed non-Abelian
gerbe can be generated in terms of a nilpotent BRST operator,
which is here constructed. The appearing fields find a natural
interpretation in terms of the universal gerbe, a 
generalisation of the universal bundle. We comment on the construction of observables in the arising Topological Quantum Field Theory. It is also shown how the BRST operator and the trace part of a suitably truncated set of fields on the non-Abelian gerbe reduce directly to the coboundary operator and the pertinent cochains of the underlying  \v{C}ech-de Rham complex.

\vfill

\begin{tabbing}
{\em Keywords}  \hspace{3mm} \= Non-Abelian gerbe, BRST symmetry,
Universal gerbe\\
{\em E-mail}    \> {\tt j.kalkkinen@imperial.ac.uk}
\end{tabbing}

\end{abstract}
\end{titlepage}

\tableofcontents
\pagebreak

\section{Introduction}
\label{Intro}
   
The natural generalisation of a principal bundle is a non-Abelian gerbe \cite{Giraud}. There are different ways of defining such an object. In this paper we shall make use of the very general approach of Ref.~\cite{BM2} based on category theory \cite{BreenAst}. Other approaches include \cite{Aschieri:2003mw,Laurent-Gengoux:2005vy}. Generalisations of Yang-Mills theory have been discussed generally \eg in \cite{Baez:2002jn,Mathai:2005sw}. Non-Abelian two-forms and their uses in loop space have been approached \eg in \cite{Hofman:2002ey,Girelli:2003ev,Baez:2004in}. An example in Supergravity can be found in \cite{Kalkkinen:1999uz}. Gerbes have appeared in String Theory \eg in \cite{Freed:1999vc,Sharpe:2000ki,Keurentjes:2001sr,Mathai:2003jx,Gawedzki:2004tu,Schreiber:2004ma}, in M-theory in \cite{Sharpe:2000qt,Kalkkinen:2004hs,Aschieri:2004yz}, and in a slightly different incarnation in Quantum Field Theory in \eg \cite{Carey:1997xm}.   
 
The aim of the Paper is to define a nilpotent BRST operator on non-Abelian gerbes and to develop methods for using non-Abelian gerbes in path integral quantisation of geometrically defined field theories. From the String Theory point of view the emphasis is on the discussion of  semiclassical backgrounds rather than defining stringy holonomies. 
As is usual in Physics, we study the geometric object through fields that live on it: in the case of a principal bundle we might equip it with a connection so that representatives of its characteristic classes can be studied conveniently.  In the case of a non-Abelian gerbe we need much more data. The requisite objects were  found in \cite{BM2} making use of recently developed methods in combinatorial differential geometry \cite{BM1}. 

We shall first cast some of the results of \cite{BM2} in a form which is perhaps more immediately applicable in  physical problems. In particular, we define the BRST operator of a non-Abelian gerbe as a nilpotent Grassmann odd operator that generates its infinitesimal symmetries. Instrumental to the construction is the universal gerbe which arises as a generalisation of the universal bundle \cite{AH}. The BRST operator can be discovered  as a covariant derivative on it. Universal gerbes in a slightly different context were  discussed also in \cite{Carey:2001gi}. As the BRST operator implements a shift symmetry, it leads to a topological theory, akin to the standard Topological Yang-Mills theory \cite{Witten:1988ze,Ouvry:1988mm}. Topological Quantum Field Theory with Abelian gerbes was also discussed more abstractly for instance in \cite{Picken:2003kw}.  
  
The method of choice for describing the structure of a fully decomposed gerbe is combinatorial differential geometry \cite{BM1}. This is perhaps unfamiliar in physics literature; a brief and informal review of the basic tools is included in Sec.~\ref{Geometry}.   Most of the discussion is on the algebraic level, and we use heuristic methods, such as path integrals, only to motivate definitions.
To give a flavour of the novelties, the requisite tool-kit contains, among other things, three different ``derivatives'': 
\begin{itemize}
\item The classical Lie-algebra valued covariant exterior derivative $\d_A$ that acts on Lie-algebra valued differential forms; 
\item The combinatorial differentials $\delta_m^{(n)}$ that act on group-valued differential forms; and 
\item 
The generalisation of the \v{C}ech coboundary operator, $\partial_\lambda$.
\end{itemize}
These differentials depend characteristically on different data: here $A$ is a locally defined Lie-algebra valued one-form, $m$ a
combinatorial Lie-group valued one-form, and $\lambda$, in the
simplest case, an element of the automorphism group of the underlying
Lie-group. 

In Sec.~\ref{Construction} the infinitesimal symmetries of a fully decomposed gerbe are found, and a provisional BRST operator $\Qn$ is written down. This provisional operator fails to be nilpotent, however, when operated on one of the ghost fields.  

Sec.~\ref{Universal} is a review of the universal bundle, and its uses for defining observables in Donaldson-Witten theory. In Sec.~\ref{UG} we generalise the construction for the non-Abelian gerbe, and write down a fully nilpotent BRST operator $\qq$ for the associated Topological Quantum Field Theory. In Sec.~\ref{Comparison} we change the grading of the BRST operator, and  show that the new operator $\QQ$ reduces to $\Qn$ on-shell. 

In Sec.~\ref{Obs} we discuss defining BRST-closed functionals. Due to the intricate structure of the field content, the simplest such functionals are also BRST-exact and therefore trivial in BRST cohomology. The complications that arise in defining invariant polynomials are intimately related to the r\^ole played by the outer automorphisms of the underlying gauge symmetry group. It remains an interesting problem to calculate the cohomology of the BRST operator. We finish by showing how the trace part of the present construction produces the Abelian gerbe \cite{Bryl}, and its symmetries.

\section{Structure of a non-Abelian gerbe}
\label{Geometry}

In this section we set the stage for later constructions. We shall first recall the basic group theoretical structures behind a non-Abelian gerbe \cite{Breen,BreenAst}, then review aspects of  differential calculus with group-valued forms \cite{BM1}, and finally summarise in Sec.~\ref{Fields} the differential geometry of a fully decomposed gerbe \cite{BM2}.

\subsection{Cohomology of a gerbe}
\label{Coho}

We give here a brief account of cohomology of gerbes. Note that this cohomology is not related {\em a priori}  to the cohomology of the BRST operator that is the main topic of this paper.

In the Abelian case, the cohomology class of a gerbe with connection and curving is a class in \v{C}ech-de Rham cohomology  \cite{Bryl}. This generalises readily to arbitrary degree. There is a well-defined characteristic class, which for an $n$-gerbe on a manifold $X$ is an element of $H^{n+2}(X, \Rset)$.

In the non-Abelian case the situation is directly analogous only at degree one: the cohomology class of a principal $G$-bundle --- seen as a zero-gerbe --- is an element of $H^{1}(X,G)$. The definition makes sense, as the cocycle condition $\lambda_{ij}\lambda_{jk}\lambda_{ki} = \unit_{i}$ is invariant  under redefinitions of the local frame $\lambda_{ij} \longrightarrow h_{i} \lambda_{ij} h^{-1}_{j}$. When the principal bundle is equipped with a connection, characteristic classes can be defined as elements of $H^{*}(X, \Rset)$ \eg in terms of invariant polynomials of the curvature of the connection.

For a one-gerbe, the cocycle condition involves both the automorphism-valued transition function $\lambda_{ij} \in \Hom({G_{j}, G_{i}})$ and the group-valued generalisation $g_{ijk} \in G_{i}$ of the Abelian \v{C}ech-cocycle
\be
\lambda_{ij}(g_{jkl}) g_{ijl} &=&g_{ijk}g_{ikl}   \\
\iota_{g_{ijk}}\lambda_{ik} &=& \lambda_{ij}\lambda_{jk} ~.
\ee
The groups $G_{i}$ could be the same group on each chart ${\cal U}_{i}$, in which case the structure is called a $G$-gerbe. In what follows we concentrate in the interest of notational simplicity on this case although the analysis goes directly over to the general $\{ G_{i} \}$-gerbe. In any case, the automorphisms $\lambda_{ij}$ are required to be invertible.

The cohomology of a non-Abelian $G$-gerbe involves, therefore, both the group $G$ and the automorphisms $\Aut G$.
Inner automorphisms $\Inn G$ are given by conjugation with a group element; outer automorphisms are the rest
\be
\Out G \eq \Aut G / \Inn G ~. \label{defOut}
\ee
For a connected, simply connected simple Lie-group $G$, $\Out G$ is given by the symmetries of the Dynkin diagram. Together with the centre of the group $\Z G$ all these groups fit in the exact sequence
\be
1\longrightarrow \Z G \longrightarrow G
\stackrel{\iota}{\longrightarrow} \Aut G
\stackrel{\sigma}{\longrightarrow} \Out G \longrightarrow 1~, \label{exsec}
\ee
and in the commutative diagram
\be
\begin{CD}
1 @>>> \Z G @>>> G @>>> \Inn G @>>> 1\\
&& @V{\iota}VV  @VVV @VVV & \\
1 @>>> 1  @>>> \Aut G  @= \Aut G @>>> 1
\end{CD}  ~.
\ee
It turns out to be useful to look upon this as a sequence of complexes $(\Z G \longrightarrow 1)$, $(G \stackrel{\iota}{\longrightarrow} \Aut G)$, and $(\Inn G {\longrightarrow} \Aut G)$.
The last column is by (\ref{defOut}) equivalent to $\Out G$, and the essence of these diagrams can be boiled down to the ``distinguished triangle'' \cite{BreenAst}
\be
\Z G[1] \longrightarrow ( G \stackrel{\iota}{\longrightarrow} \Aut G ) \longrightarrow  \Out G ~,
\ee
where ``$[1]$'' indicates the shift in degree.

More technically this can be summarised by saying that the cohomology class $(\lambda_{ij},g_{ijk})$ of a non-Abelian $G$-gerbe is valued in the crossed module $G \stackrel{\iota}{\longrightarrow} \Aut G $, denoted here with $\mathbf{G}$. The group $H^{1}(X, \mathbf{G})$ of equivalence classes of such gerbes fits in the exact sequence \cite{Breen} 
\be
H^{0}(X, \Out G) \longrightarrow H^{2}(X, \Z G) \longrightarrow H^{1}(X, \mathbf{G}) \longrightarrow \Tors(\Out  G) ~,
\ee
where $\Tors H$ refers to isomorphism classes of principal $H$-bundles. The shift in the degree is due to the the fact that $\mathbf{G}$ is a complex. Therefore, if there are no outer automorphisms, the gerbe $\mathbf{G}$ is cohomologically an Abelian $\Z G$-gerbe.

\subsection{Group-valued differential forms}
\label{Forms}

In this section we review informally basic techniques for calculating with group-valued differential forms needed later in the paper. For a more systematic account, see \cite{BM1,BM2}.

Let $\OMEGA^*(X,G)$ denote the sheaf\footnote{For a precise definition see \cite{BM1}.} of group $G$-valued local differential forms on $X$ relative to a fixed cover $\{ {\cal U}_{i} \} $ of $X$. To be quite concrete, a typical element in it is a rank-$n$  combinatorial differential form $\alpha_{i_{1}\cdots i_{k}}$   in $\Omega^{n}( {\cal U}_{i_{1}\cdots i_{k}},G_{i_{1}})$ defined on the $k$-fold intersection ${\cal U}_{i_{1}\cdots i_{k}}$ with coefficients in the local group $G_{i_{1}}$. This generalises the
\v{C}ech-de Rham complex in a natural way. In what follows there is no need to indicate explicitly what the local group $G_{i_{1}}$ is, because it is implicit in the first index of the intersection, $i_{1}$: in lieu of $G_{i_{1}}$ we write simply $G$.

Let $g,h \in \OMEGA^*(X,G)$ and $\mu,\nu,\lambda  \in
\OMEGA^*(X,\Aut(G))$. The group commutator for $G$ and $\Aut G$-valued fields is defined as \be [g,h] &\eq& ghg^{-1}h^{-1} ~, \ee
and similarly for $\Aut G$. When the degree of both fields is
positive, the group commutator reduces (on the level of  one-jets
\cf \cite{BM1}) to the classical Lie-bracket \be [g,h] &=& gh - hg
~. \ee This is simply because on the level of one-jets $g = 1 + x
+ {\cal O}^{2}$ and  $h = 1 + y + {\cal O}^{2}$ so that $[g,h] =
xy -yx + {\cal O}^{3}$. In combinatorial differential calculus it
is unnecessary to distinguish notationally between, for instance,
$g$ and $x$, and we shall indeed  change the point of view from the
group level to the algebra level as suitable.

The action of elements $\mu$ of $\OMEGA^*(X,\Aut(G))$ on those of
$g \in \OMEGA^*(X,G)$ is denoted as $\mu(g)$. Their commutator is
defined as \be [\mu, g] & \eq & \mu(g) g^{-1} \qquad  \in
\OMEGA^*(X,G)~, \ee which on one-jet level $\OMEGA^*(X,\Lie(G))$
reduces for positive degree fields to the classical (graded) Lie
bracket.  One can also define the bracket \be [g, \mu] &\eq&
-[\mu, g] ~, \ee which is still group-valued. When the degree of
both fields is again positive, it too reduces to classical Lie
brackets.

All of these classical Lie-brackets, be their arguments $A,B,C$
$\Lie G$ or $\Lie \Aut G$-valued differential forms, obey  the
usual graded classical Jacobi identity, \be (-)^{|A||C|} [ A , [
B, C]] + (-)^{|C||B|} [ C , [A , B]] + (-)^{ |B||A|} [ B , [ C,
A]] = 0 \ee as well as the Leibnitz rule \be \d_{m} [A,B] &=& [
\d_{m}A, B] + (-)^{|A|} [A, \d_{m}B] ~, \ee where $|A|$ is the
degree of $A$ \etc, and $\d_{m}$ is a classical covariant
derivative \be \d_{m} & \eq & \d + [m, ~ \cdot ~] ~. \ee

\begin{definition}
\label{DefGroup} The adjoint action $\iota : G \longrightarrow
\Aut(G)$ is
\be
\iota_g(h) &\eq& g h g^{-1} ~.
\ee
The automorphism group $\Aut G$ acts on itself by
\be
{}^{\lambda}\nu &\eq& {\lambda} \nu {\lambda}^{-1} ~.
\ee
\end{definition}

\begin{lemma}
Adjoint action $\iota_{g}$ by a group element $g \in G$ enjoys the
properties
\be
[g,h] &=& [\iota_g, h] \\
{}^\lambda \iota_g &=& \iota_{\lambda(g)} \\
{} [\lambda, \iota_g] &=& \iota_{[\lambda, g]} ~.
\ee
where $\lambda \in \Aut G$ and $h \in G$.
\end{lemma}
\begin{proof}
First, definition of the commutators $[g,h] = gh g^{-1}h^{-1} =
\iota_g(h) \, h^{-1}  =  [\iota_g, h]$; second, elements of
$\Aut(G)$
are homomorphisms and ${}^\lambda\mu = \lambda\mu \lambda^{-1}$;
third, $[\lambda,\iota_g](h) = \lambda(g) \, g^{-1}hg \,
\lambda(g)^{-1} = [\lambda,g] h [\lambda,g]^{-1}$.
\end{proof}

The combinatorial covariant derivatives
\be
\delta_{m}^{{(n)}} :
\OMEGA^{n}(X,G) \longrightarrow \OMEGA^{n+1}(X,G) ~, \qquad n \geq
0
\ee
reduce to the classical covariant derivatives
\be
\delta_{m}^{(n)} \omega &=& \d \omega + [m,\omega]  =  \d_{m}
\omega , \qquad n \geq 2 \\ \delta_{m}^{(1)} \omega &=& \d_{m}
\omega + \half [\omega, \omega] ~,
\ee
except for $n=0$. Note that
$\delta_{m}^{(n+1)}\delta_{m}^{(n)} \omega = [\kappa(m), \omega]$
for $n=0$ and $n \geq 2$ with
\be
\kappa(m) &=& \d m + \half [m,m]
~, 
\ee
whereas for $n=1$ we have
\be
\delta_{m}^{(2)}\delta_{m}^{(1)} \omega = [\kappa(m) + \d_m
\omega, \omega]~.
\ee
Also
\be
\delta_{m}^{(1)} (-\omega) &=& -
\delta_{m}^{(1)} \omega +  [\omega, \omega] ~.
\ee
There is an
alternative set of differentials
\be
\tilde \delta_{m}^{(n)}
\omega &\eq& - \delta_{m}^{(n)} (-\omega)
\ee
which of course
coincides the with $\delta_{m}^{(n)}$ for $n \geq 2$. The analogue of the \v{C}ech differential is
\be
\partial_{\lambda} \omega_{ij} & \eq &  \omega_{ij} + \lambda_{ij}(\omega_{jk}) +
\lambda_{ij}\lambda_{jk}(\omega_{ki})
~.
\ee

\subsection{A fully decomposed gerbe}
\label{Fields}

The differential geometry of a non-Abelian gerbe \cite{BM2}
involves the fields summarised in Table \ref{fieldsTable1}.
\begin{table}
\be
\begin{array}{|c|cccc|}
  \hline
    & \text{0-form} & \text{1-form} & \text{2-form} & \text{3-form}
    \\
  \hline
  G & g_{ijk} & \gamma_{ij}  & B_i, ~\delta_{ij} & \omega_i \\
  \Aut(G) & \lambda_{ij} & m_i & \qquad   \nu_i &   \\
  \hline
\end{array}
\nonumber
\ee
\mycaption{The local fields on a non-Abelian gerbe.
\label{fieldsTable1}}
\end{table}
The {\em cocycle  data} $(\lambda_{ij}, g_{ijk})$ satisfies
\be
\lambda_{ij}(g_{jkl}) g_{ijl} &=&g_{ijk}g_{ikl} \label{funco1} \\
\iota_{g_{ijk}}\lambda_{ik} &=& \lambda_{ij}\lambda_{jk} ~.
\label{funco2}
\ee
We add to this the connection $m_i$ and the two-form $B_i$ and
define
\be
\omega_i &\eq&\d_{m_i}(B_{i}) \label{omegaDef} \\
\nu_i&\eq& \kappa(m_i) - \iota_{B_i} ~. \label{nuDef}
\ee
The covariant derivative is the standard $\d_{m} B \eq \d B + [m,
B]$ with curvature $\kappa(m) \eq \d m + \half [m,m]$. It is
compatible with the inner action $\iota$  in the sense that
$\iota_{d_{m}(B)} = {d_{m}\iota_B}$. Note also that we will use
this definition inherited from Lie-algebra valued differential
forms everywhere, including in the case of one-forms where
Refs.~\cite{BM1,BM2} use\footnote{The factor of $1/2$ is crucial:
this definition together with the result (6.1.19) of \cite{BM2}
$\delta^1(-\gamma_{ij}) =
-\delta^1(\gamma_{ij})+[\gamma_{ij},\gamma_{ij}]$ can be used to
turn the cocycle condition (6.1.18) of \cite{BM2} $
\delta^1(-\gamma_{ij}) =
-\d_{m_i}(\gamma_{ij})+(1-\half)[\gamma_{ij},\gamma_{ij}]_{m_{i}}$
into the form required here in Eq.~(\ref{deltaDeff}).}
$\delta^{1}_{m}(\gamma) \eq \d_{m} \gamma + \half
[\gamma,\gamma]_{m}$.

To relate these fields $m_{i}$ and $B_{i}$ on different charts, we
need $\gamma_{ij}$
and $\delta_{ij}$ such that
\be
{}^{\lambda_{ij}*}m_j -m_i &=& -\iota_{\gamma_{ij}} \label{gammaDef}
\\
\lambda_{ij}(B_j) - B_i &=& \delta_{ij} - \d_{m_i}(\gamma_{ij})+
\half [\gamma_{ij}, \gamma_{ij}]_{m_i} ~. \label{deltaDef}
\ee
The star in the action of $\lambda_{ij}$ here refers to the fact
that $m_{i}$ transforms as a gauge field ${}^{\lambda_{ij}*}m_j
\eq {}^{\lambda_{ij}}m_j + {\lambda_{ij}} \d {\lambda_{ij}}^{-1}$.
We can view (\ref{deltaDef}) as a definition of the $G$-valued
two-form $\delta_{ij}$
\be
 \delta_{ij} &\eq& \lambda_{ij}(B_j) - B_i + \d_{m_i}(\gamma_{ij})
 -\half  [\gamma_{ij}, \gamma_{ij}]_{m_i},\label{deltaDeff}
\ee
whereas (\ref{gammaDef}) determines only the inner action of
$\gamma_{ij}$. Note that the twisted commutators $[\gamma_{ij},
\gamma_{ij}]_{m_i}$ are actually independent of the twisting one-form
$m_{i}$, \cf (A.1.23) of \cite{BM2}, so we can treat them
safely as standard untwisted commutators. This leads
\cite{BM2} to the
cocycle conditions
\be
\partial_{\lambda_{ij}}(\gamma_{ij}) &=& \tilde\d_{m_i}(g_{ijk})
\label{gammaDiff}  \\
\partial_{\lambda_{ij}}(\delta_{ij}) &=& [\nu_i, g_{ijk}] ~.
\label{deltaDiff}
\ee
The covariant derivative of group-valued functions
$\tilde\d_{m_i}(g_{ijk})$ cannot be easily represented in terms of
algebra-valued expressions.\footnote{The notation of
\cite{BM2} used $\tilde\delta^0(g_{ijk})  =
{}^{g}\delta^{0}_{m_{i}}(g_{ijk}) = \tilde\d_{m_i}(g_{ijk})$.}

We call the triple  $(m_i,\gamma_{ij},B_i)$ {\em connection data}.
Here $\delta_{ij}$ and $\nu_i$ belong to the {\em curvature
triple} $(\nu_i,\delta_{ij},\omega_i)$. The cocycle conditions and
the transformation properties of the curvature triple are in
addition to the above equations
\be
\iota_{\omega_i} &=& - \d_{m_i}(\nu_i) \label{k1} \\
\d_{m_i}(\omega_i) &=& [\nu_i, B_i] \label{k2} \\
{}^{\lambda_{ij}}\nu_j -\nu_i &=& - \iota_{\delta_{ij}} \label{k3}\\
\lambda_{ij}(\omega_j) - \omega_i &=& \d_{m_i}(\delta_{ij}) +
[\gamma_{ij},\nu_i] - [\gamma_{ij},\delta_{ij}] \label{k4}
\ee
One of the consequences of these cocycle conditions is the fact that if the {\em fake curvature} $\nu_{i}$ vanishes, then by (\ref{k1}) and  (\ref{k3}) the rest of the curvature data are Abelian.

\subsection{Exact symmetries}
\label{Exact}

The freedom to choose a basis in each chart ${\cal U}_{i}$ of the
manifold gives rise to the {\em local gauge symmetry}: given local
functions $h_i \in \Omega^0({\cal U}_{i},\Aut(G_i))$ we may change the basis by
\be
m_{i} &\longrightarrow & {}^{h_{i}*}m_{i} \qquad \eq h_{i}
\d_{m_{i}}(h_{i}^{-1}) \\
\gamma_{ij} &\longrightarrow & h_{i} (\gamma_{ij}) \\
B_{i}&\longrightarrow & h_{i}(B_{i})
\ee
and so on. Under these symmetries the cocycle conditions transform
obviously covariantly. Connection data deserves its name because it
can be shifted by {\em affine data}
$(\pi_i,\eta_{ij},\alpha_i,E_{i})$ that satisfy the cocycle
conditions \cite{BM2}
\be
{}^{\lambda_{ij}}\pi_j -\pi_i &=& -\iota_{\eta_{ij}} \label{piCos}\\
\partial_{\lambda_{ij}}(\eta_{ij}) &=& [\pi_i, g_{ijk}] ~.
\label{etaCos}
\ee
The transformation rules of the connection data are
\be
m_i'-m_i &=& \pi_i \quad + \iota_{E_i} \label{mAff} \\
\gamma_{ij}'-\gamma_{ij} &=& \eta_{ij}  \quad - \lambda_{ij}(E_j)
+ E_i \label{gammaAff}
\\
B_i'-B_i &=& \alpha_i  \quad +  \kappa({E_i}) + [m_i,E_i] +
[\pi_i,E_i] ~. \label{BAff}
\ee
This induces the following symmetry on the curvature triple:
\be
\nu_i' - \nu_i &=& \kappa(\pi_i) +
[m_i,\pi_i]  - \iota_{\alpha_i} \label{nuAff} \\
\delta_{ij}' - \delta_{ij} &=&   \lambda_{ij}(\alpha_j)
- \alpha_i  + \d_{m_i}(\eta_{ij}) - [\eta_{ij},\eta_{ij}]_{m_i}
\nonumber \\
& &  + [\pi_{i},\eta_{ij}]_{m_i} - [\gamma_{ij},\eta_{ij}]_{m_i}
+ [\gamma_{ij},\pi_{i}]_{m_i} \label{deltaAff} \\
\omega_i' - \omega_i &=& \d \alpha_i + [m_i,\alpha_i] + [\pi_i,B_i
+ \alpha_i]  \nonumber \\
& &  - [ \alpha_i,E_i] + [\nu_i +  \kappa(\pi_i) + [m_i,\pi_i]
,E_i] \label{omegaAff}
\ee
We call this symmetry the {\em affine gauge symmetry}.
The affine data are themselves subject to the symmetry
\be
\pi' -\pi &=& \iota_{\rho_i} \\
\eta_{ij}' - \eta_{ij} &=& \rho_i - \lambda_{ij}(\rho_i)  \\
E_i' - E_i &=& - \rho_i \\
\alpha_i'-\alpha_i &=&   \kappa({\rho_i}) + [m_i,\rho_i] +
[\pi_i,\rho_i] ~. \label{rhoAff}
\ee
We call this redundancy the {\em reduced gauge symmetry}.

\section{Infinitesimal symmetries}
\label{Construction}

A BRST operator is a nilpotent (of order two) differential  on the
field space.\footnote{See \cite{Henneaux:1992ig} for a thorough
treatment. We use here the concept of ``field space''
heuristically; in Sec.~\ref{UG} we present a more detailed
description of what we mean by it.} In order to construct such an
operator, we must be able to model differentials of physical
fields consistently. We do this formally using the Grassmann
algebra $\Gset$ of anticommuting real numbers. The infinitesimal
fields are called ghosts in the physics literature, as they
decouple from physical amplitudes. The requirement for nilpotency
of the BRST operator may require introducing differentials for
ghost fields themselves as well; these objects are called
ghost-for-ghost fields. Ghosts-for-ghosts can be thought of as
two-forms in the field space. All the emerging fields are graded
in terms of the ghost number, and there can, in principle, be an
infinite tower of them, though we shall here have to advance 
up to ghost number three only.

As a field with positive ghost number is an infinitesimal,  it
gets  also at form degree zero its values in the Lie-algebrae
$\Lie G$ and $\Lie G\otimes \Gset$ rather than the respective
groups.  Indeed, a typical field of odd ghost number is a
differential form in $\OMEGA^{n}(X, \Lie G \otimes \Gset)$ for
some $n>0$; for even positive ghost number they are classical differential
forms in  $\OMEGA^{*}(X, \Lie G)$. This potential discrepancy with 
combinatorial differential forms will be explained and put in context  in Sec.~\ref{UG}.

In this section we shall begin by writing down a BRST  operator
``$\qqq$'' that generates infinitesimal versions of the gauge
transformations of the last section. Reducibility and nilpotency considerations force us to amend the derivative $\qqq$ to ${\Qn} = \qqq + \delta + \tilde\delta$. The BRST operator we obtain in this way is nilpotent on connection data, but fails to be nilpotent on one of the ghost fields.

\subsection{Infinitesimal transformations}
\label{Infinites}

The derivative ``$\qqq$'' arises from infinitesimal
displacements generated by local gauge transformations $h_{i}$ and
the symmetries of the gerbe in Sec.~\ref{Exact}. For the finite
local gauge transformation $h_{i} \in \Omega^{0}({\cal U}_{i},\Aut G)$
corresponds the infinitesimal, Grassmann-valued ghost field $c_{i}
\in \Omega^{0}({\cal U}_{i},\Lie \Aut G \otimes \Gset)$. Similarly, the
affine data $(\pi_i,\eta_{ij},\alpha_i,E_{i})$ of Sec.~\ref{Exact}
are all Grassmann-valued ghost fields in this section. We may now
write down the local gauge and affine transformations {\em in
infinitesimal form} for the gauge fields
\be
\qqq m_i &=& \pi_i + i_{E_i} -\d_{m_i}c_i \\
\qqq \gamma_{ij} &=& \eta_{ij} - \lambda_{ij}(E_j) + E_i + [c_i,
\gamma_{ij}] \\
\qqq B_i &=& \alpha_i + \d_{m_i}(E_i) + [c_i, B_i] \\
\qqq \alpha_i &=&  - [\pi_i, E_i] + [ c_i, \alpha_i] \label{ekstra} \\
\qqq c_i &=& \half [c_i, c_i] ~.
\ee
Other fields $x$ transform according to the standard rule
\be
\qqq x = [c_{i}, x] ~.
\ee
As the cocycle data remain constant under the symmetries of the
gerbe we set (\cf Sec.~\ref{CovDer})
\be
\qqq \lambda_{ij} &=& 0 \label{Ql} \\
\qqq g_{ijk} &=& 0 ~.
\ee

The connection data have ghost number zero, and all the
transformation parameters above
$c_{i},\pi_i,\eta_{ij},\alpha_i,E_{i}$ have ghost number one. The derivative $s$ raises ghost number by one. Ghost
number grading is independent of form degree grading. In this
section, a field of form degree $p$ and ghost number   $q$ can be
thought of as a real number-valued differential form of degree $p$
when $q$ is {\em even}. When $q$ is {\em odd}, the components of
the differential form are elements of the Grassmann algebra
$\Gset$. By multiplying two such objects of grading $(p,q)$ and
$(p',q')$ we get an object of grading $(p+p', q+q')$. These two
bigraded objects are mutually odd precisely when $p p' + q q'$ is
odd, otherwise even.\footnote{There is an other way of doing this,
\cf Sec.~\ref{Grading}. This convention is the only one
immediately  consistent with the infinitesimal symmetries of the
gerbe though.} The brackets $[~,~]$ can be graded so that the
pertinent graded Jacobi identity applies.

The differences to the original transformations in
Sec.~\ref{Fields} are the following:
\begin{itemize}
\item
We have discarded terms  that are of higher order than linear in
affine data, namely $[E_i,E_i]/2 + [\pi_i,E_i]$ in  $\qqq B_i$
(\ref{BAff}),  and $[\rho_i,\rho_i]/2 + [\pi_i,\rho_i]$ in  $\qqq
\alpha_i$ (\ref{rhoAff}).
\item
We have added the extra term $ - [\pi_i, E_i]$ in (\ref{ekstra}).
Note that the same term had to be struck off from (\ref{BAff}).
Also, this term vanishes at the equivariant $c_{i} = 0$ fixed
point locus  of the full nilpotent BRST operator, \cf
Sec.~\ref{Elim}.
\item
Commutators between two affine-fields-turned-ghosts have been
changed to anticommutators, and vice versa.
\end{itemize}
The justification for these differences is the fact that $\qqq$
does still generate symmetries of the underlying gerbe, though
infinitesimal.

The fact that $\qqq x$ is infinitesimal of order one means that we
can extend the action of $\qqq$ to any functional composed of
fields whose  BRST transformation under $\qqq$ we have defined.
$\qqq$ is therefore a graded odd derivation, and raises ghost
number by one. Most importantly, it is nilpotent
\be
\qqq^{2} = 0 ~.
\ee

\subsection{Reducibility}
\label{Redu}

All of the fields $\pi_{i}$, $E_{i}$, and $c_{i}$ describe shifts
in $m_{i}$ in different ways. Given a specific, fixed shift $m'_i
- m_i$ there is latitude in how it is written down in terms of
$\pi_{i}$, $E_{i}$, and $c_{i}$. In Sec.~\ref{Exact} the latitude
in the choice of $(\pi_{i}, E_{i})$ was parametrised in terms of
$\rho_{i}$. Taking also $c_i$ into account we need two more
ghost-for-ghosts $\varphi_{i} \in \Omega^{0}({\cal U}_{i},\Lie \Aut G)$ and
$\phi_{i} \in \Omega^{0}({\cal U}_{i},\Lie G)$.

The new fields force us to amend the BRST differential  $\qqq
\longrightarrow \qqq + \delta$. The new piece, $\delta$, is an odd
graded derivation of ghost number one, as was $\qqq$. The nontrivial
action of $\delta$ is
\be
\delta \pi_{i} &=& \d_{m_{i}} \varphi_{i} + \iota_{\rho_{i}} \\
\delta E_{i} &=&  \d_{m_{i}} \phi_{i} - {\rho_{i}} \\
\delta c_{i} &=& \varphi_{i} + \iota_{\phi_{i}} \\
\delta \alpha_{i} &=& \d_{m_{i}} \rho_{i} + [B_{i}, \varphi_{i}] + [
\phi_{i},  \nu_{i}]  \\
\delta \eta_{ij} &=& \rho_{i}-d_{m_{i} }\phi_{i}
-\lambda_{ij}\Big(\rho_{j}-\d_{m_{j}}\phi_{j}\Big)  +
[\gamma_{ij},\phi_{i}] + [ \gamma_{ij},\varphi_{i}] ~. \label{deta}
\ee
These transformations are chosen so that $\delta \qqq m_{i}  =
\delta \qqq \gamma_{ij}  =  \delta \qqq B_{i}  =  0$. As the
action of  $\delta$ on other fields is trivial, $\delta$ is
nilpotent $\delta^{2}  =  0$. Note that $\qqq + \delta$ is not
nilpotent, but, for instance,
\be
(\qqq + \delta)^{2} E_{i} &=& [\varphi_{i},E_{i}] + [\pi_{i},\phi_{i}] ~.
\ee

This non-nilpotency can be remedied partially by taking into
account that there is a  further latitude in defining the
$\rho_{i},\phi_{i},\varphi_{i}$ system. This latitude has to be
parametrised  with the ghost number-three field $\sigma_{i} \in
\Omega^{0}({\cal U}_{i},\Lie G \otimes \mathbb{G})$. This gives rise to the transformations
\be
\bar\delta \varphi_{i} &=& -\iota_{\sigma_{i}} \\
\bar\delta \phi_{i}&=& \sigma_{i}\\
\bar\delta \rho_{i}&=&  \d_{m_{i}} \sigma_{i} + [ \varphi_{i}
,E_{i}]  - [\phi_{i}, \pi_{i}]  \\
\bar\delta \sigma_{i}&=&   [\phi_{i}, \varphi_{i}] ~.
\ee
The construction is such that $\bar\delta \delta (\pi, E, c) =0$.
Again $\bar\delta$ annihilates all other fields so that
$\bar\delta^{2}  =  0$.

\begin{theorem}
\label{fullQ} The operator ${\Qn} \eq \qqq + \delta + \bar\delta$
is  an odd derivation of ghost number one. It is nilpotent
$\Qn^{2} x=0$ on all fields $x$ where we have defined it, except
on $x=\eta_{ij}$
\be
{\Qn}^{2} \eta_{ij}
&=& -\Big[
{}^{\lambda_{ij}}\Big(\varphi_{j} + \iota_{\phi_{j}}\Big) - (\varphi_{i} + \iota_{\phi_{i}}) , ~~~ \lambda_{ij}(E_{j})\Big] \nonumber \\
&& - \Big[{}^{\lambda_{ij}}c_{j} - c_{i} , ~~~ {\lambda_{ij}}(\rho_{j}-\d_{m_{j}}\phi_{j}) \Big]
~.
\ee
\end{theorem}
The operator ${\Qn}$ does therefore not quite qualify as  a BRST
operator. Note that ${\Qn}$ is nevertheless nilpotent in
particular on the connection data
\be
{\Qn}^{2} B_{i} = {\Qn}^{2} \gamma_{ij}
= {\Qn}^{2} m_{i} = 0 ~.
\ee
Furthermore, the above obstruction to nilpotency vanishes if
$\varphi_{i} + \iota_{\phi_{i}}$ and $c_{i}$ extend to global  sections.

The resolution to this problem is to introduce new fields $a_{ij}$
and $b_{ij}$ that control the behaviour of $c_{i}$ and
$\varphi_{i}$ on double-intersections ${\cal U}_{ij}$. In pursuing
this, the relationship of the BRST operator to the symmetries of
the underlying gerbe becomes slightly obscured. This happens
necessarily because the procedure requires essentially replacing,
for instance, the term  $d_{m_{i}} \phi_{i}
-\lambda_{ij}(\d_{m_{j}}\phi_{j})$ in (\ref{deta}) by the
covariant derivative of one of the new fields $\d_{m_{i}}b_{ij}$.

Instead of amending $\Qn$ here further, in Sec.~\ref{UG} we will
define an operator $\qq$ which is  by construction nilpotent, and
that  reduces  to $\Qn$ on-shell. This will require a more
geometric understanding of the BRST differential at our disposal.

\subsection{The curvature triple}
\label{Triple}

To each (field, ghost) pair one may associate a curvature as in
Table \ref{strengthsTable}.
\begin{table}
\be
\begin{array}{|cccc|}
\hline
(\mathrm{field}, \mathrm{ghost} ) &   & \mathrm{curvature} &
\mathrm{domain}\\
\hline
(m_i, \pi_i) & \longrightarrow & \nu_i & {\cal U}_{i} \\
(\gamma_{ij}, \eta_{ij}) & \longrightarrow  & \delta_{ij} & {\cal
U}_{ij} \\
(B_i, \alpha_i) & \longrightarrow  & \omega_i & {\cal U}_{i} \\
\hline
\end{array}
\nonumber
\ee
\mycaption{Fields and their field strengths.
\label{strengthsTable}}
\end{table}
It can be shown now that to linear order the local, affine, and
reduced gauge transformations of the curvature triple
$(\nu_i,\delta_{ij},\omega_i)$ in Sec.~\ref{Exact} arise, as
expected, from those of the underlying connection data modulo
terms that vanish when the constraints
\be
\C_{ij}^{0} &\eq& {}^{\lambda_{ij}*}m_j -m_i + \iota_{\gamma_{ij}}
\label{c1} \approx 0 \\
\C_{ijk}^{0} &\eq& \partial_{\lambda_{ij}}(\gamma_{ij})-\tilde\d_{m_i}(g_{ijk})
\approx 0 \label{c2} \\
\B_{ij}^{1} &\eq& {}^{\lambda_{ij}}\pi_j -\pi_i +\iota_{\eta_{ij}}
\approx 0 \label{c3}  \\
\B_{ijk}^{1} &\eq& \partial_{\lambda_{ij}}(\eta_{ij}) - [\pi_i,
g_{ijk}]\approx 0  \label{c4}
\ee
are imposed. These constraints arose as cocycle conditions in
(\ref{gammaDef}), (\ref{gammaDiff}), (\ref{piCos}),  and
(\ref{etaCos}); their r\^ole is to relate data on different charts
to each other.

The cocycle conditions (\ref{deltaDiff}) and (\ref{k1}) --
(\ref{k4}) for the curvature triple  are similarly satisfied up to
terms proportional to these constraints. The cocycle conditions
(\ref{k1}) -- (\ref{k4}) are easy enough to verify  using standard
de Rham calculus with  Lie-algebra valued differential forms. The
cocycle condition (\ref{deltaDiff}) requires special attention,
however,  because it involves derivatives of a group-valued local
{\em function}. We explain how this comes about carefully in
\begin{theorem}
\label{deltaTheo}
The cocycle condition
\be
\partial_{\lambda_{ij}} \delta_{ij} &\approx& [\nu_{i}, g_{ijk}]
\ee
arises as a consequence of the cocycle conditions (\ref{gammaDef})
and (\ref{gammaDiff}) or, equivalently, as a consequence of the
constraints $\C_{ij}^{0} \approx\C_{ijk}^{0} \approx 0$.
\end{theorem}
\begin{proof}
One applies  $\partial_{\lambda_{ij}}$ on the definition of $\delta_{ij}$
in Eq.~(\ref{deltaDeff}). Note first that the derivative
$\partial_{\lambda_{ij}}$ has the essential property
\be
\partial_{\lambda_{ij}}\Big( \lambda_{ij}(B_j)-B_i \Big) &=&
- [\iota_{B_i}, {g_{ijk}}] ~.
\ee
After obvious cancellations this can be seen as follows:
\be
\iota_{g_{ijk}}(B_i) - B_i = {g_{ijk}}B_i {g_{ijk}}^{-1} \,  B_i^{-1}
= [{g_{ijk}}, B_i] = - [B_i, {g_{ijk}}] = - [\iota_{B_i}, {g_{ijk}}]
~.
\ee
Using now repeatedly the constraint $\C_{ij} \approx 0$
we arrive at the expression
\be
\partial_{\lambda_{ij}} \delta_{ij} &\approx&  - [\iota_{B_{i}}, g_{ijk}]
+\d_{m_{i}} \partial_{\lambda_{ij}} \gamma_{ij} -  \half [
\partial_{\lambda_{ij}} \gamma_{ij} ,  \partial_{\lambda_{ij}} \gamma_{ij} ]  ~.
\ee
At this point we have to return to the group-valued differential
forms, and the notation of Ref.~\cite{BM2}: the above-appearing expression
\be
\d_{m_{i}} \partial_{\lambda_{ij}} \gamma_{ij} -  \half [ \partial_{\lambda_{ij}}
\gamma_{ij} ,  \partial_{\lambda_{ij}} \gamma_{ij} ]
\ee
can then be cast in the form
\be
= &
\delta^{1}_{m_{i}} (\partial_{\lambda_{ij}} \gamma_{ij}) -
[\partial_{\lambda_{ij}} \gamma_{ij},~ \partial_{\lambda_{ij}} \gamma_{ij}]
& \qquad \text{ by def.~of $\delta^{1}_{m}$} \\
= &
- \delta^{1}_{m_{i}} (-\partial_{\lambda_{ij}} \gamma_{ij})
& \qquad \text{ by Eq.~(6.1.19) of \cite{BM2}} \\
\approx &
- \delta^{1}_{m_{i}} (-\tilde\delta^{0}_{m_{i}} g_{ijk})
& \qquad \text{ by Eq.~(\ref{gammaDiff})} \\
= &
- \delta^{1}_{m_{i}} (\delta^{0}_{m_{i}} (-g_{ijk}))
& \qquad \text{ by Remark 6.1 of \cite{BM2} }\\
= &
- [[ \kappa(m_{i}),~ g_{ijk}^{-1}]]
& \qquad \text{ by Eq.~(A.1.13) of \cite{BM2}} \\
= &
+ [ \kappa(m_{i}),~ g_{ijk} ] ~.
\ee
Combining this with the definition of $\nu_{i}$ concludes the proof.
\end{proof}

\section{Topological Yang-Mills Theory}
\label{Universal}

Consider isomorphism classes $p \in {\cal P}$ of principal bundles
$L_{p}$ with connection and possibly other data on a fixed
manifold $X$. The universal bundle
$\mathfrak{P} \longrightarrow X \times {\cal P}$ fits in the
diagram
\be
\begin{CD}
L_{p} @>>> \mathfrak{P} \\
@V{\pi}VV   @VV{\Pi}V \\
X \times \{ p \} @>{i}>> X \times {\cal P} ~.
\end{CD} \label{UBdiag}
\ee
In the case of Topological Yang-Mills Theory
\cite{Witten:1988ze,Atiyah:1990tm} (\cf \cite{Birmingham:1991ty}
for a review) we fix the local transition functions $\ell_{ij}$ consistently $\ell_{ij}\ell_{jk}\ell_{ki}=\unit$ once
and for all, and keep free only the local connection one-form $m \in
\A$ on the bundle. The arising universal bundle $\mathfrak{P}$ is locally
of the form $L_{p} \times \A$. The gauge equivalence classes of
the connections ${\cal P} = \A / \G$ do not necessarily form a
smooth manifold; the universal bundle, nevertheless, has a smooth
base space, which is locally of the form $(L_{p} \times \A)/ \G$.
If $\G$ acts freely on $\A$, this reduces to $X \times \A / \G$,
and we identify ${\cal P} = \A/\G$.

As all objects transform in the Yang-Mills case covariantly between charts, there is therefore no need to indicate the local chart, and we will suppress the pertinent indices for a moment. Choosing $(g_{ijk}, \lambda_{ij}) = (\unit, \ell_{ij})$ the non-Abelian gerbe collapses now to Topological Yang--Mills theory with only $m, \pi, c, \varphi$ active, and other fields set to trivial values.

Given a one-form $c \in T^{*} \A$, we may construct a covariant exterior derivative on $\mathfrak{P}$
\be
D_{\mu} &=& \d_{m} + \qq_{c} ~, 
\ee
where $\qq_{c} X = \qq X + [c,X]$. Here $\d$ is the exterior
derivative on $X$ and $\qq$ on $\A/\G$, when the  latter makes
sense. The curvature can be expanded in terms of the bidegree
\be
D_{\mu}^{2} &=& \F^{(2,0)} + \F^{(1,1)} + \F^{(0,2)}  \\
&\eq& \kappa(m) + \pi + \varphi ~;
\ee
with the latter line we merely mean that \eg the field $\pi$
stands for the $(1,1)$ component of the curvature.  These
definitions imply then, in fact, together with the standard
Bianchi identity $D_{\mu}\F =0$, the action of $q$ on various
fields
\be
\qq m &=& \pi -\d_{m}c \\
\qq c &=& \varphi - \half [c,c] \\
\qq \pi &=& -\d_{m}\varphi -[c,\pi] \\
\qq \varphi &=& -[c, \varphi] ~.
\ee

\subsection{Observables}
\label{TYMobs}

The Bianchi identity implies also
\be
(\d + \qq) \Tr \F^{n} &=& 0~.
\ee
Let us decompose $ \Tr \F^{n}  = \sum_{k} W_{k}$, where $k$ is the
form degree on $X$, and integrate each form over $\gamma \subset
X$ of suitable  dimension $k$ \cite{Witten:1988ze}
\be
W_{k}(\gamma) &\eq& \langle \int_{\gamma}W_{k} \rangle ~. \label{oob}
\ee
For this to be a good observable $\gamma$ should not have a
boundary, as only then is (\ref{oob})  BRST-closed
\be
\qq W_{k}(\gamma) &=& - \langle \int_{\gamma}\d W_{k-1} \rangle  =0 ~.
\ee
On the other hand, changing $\gamma$ by a boundary $\partial s$
changes this observable by a BRST-exact term
\be
W_{k}(\partial s) &=& \langle \int_{s} \d W_{k} \rangle  =  - \langle \qq \int_{s} W_{k-1} \rangle  = 0~,
\ee
so that the vacuum expectation value $W_{k}(\gamma)$ remains
invariant. The vacuum expectation values $W_{k}(\gamma)$ depend
therefore only on the homology class $[\gamma] \in H_{k}(X)$. In
the case of Topological Yang-Mills, $W_{k}(\gamma)$ are the
Donaldson-Witten invariants \cite{Witten:1988ze}.

\subsection{Curvature}
\label{Curva}

Let us decompose --- following Ref.~\cite{Birmingham:1991ty} and references therein --- the BRST
operator into a horizontal and vertical parts
\be
\qq &=& \qqH + \qqV ~, 
\ee
where $\qqV$ acts along the fibre ${\cal G}$, and
$\qqH$ on the base ${\cal A}/{\cal G}$.

The vertical derivative generates standard gauge transformations
\be
\qqV m &=& -\d_m c \\
\qqV c &=& - \half [c,c]\\
\qqV x &=& - [c,x] ~.
\ee
It is nilpotent $(\qqV)^2=0$, so that its curvature
vanishes identically. The horizontal part has curvature
\be
(\qqH)^2 m &=& - \d_m \varphi \\
(\qqH)^2 x &=& [\varphi, x] ~,
\ee
where $x$ is any other field than $m$. One may think of $\qqH$ as the {\em covariant exterior derivative} \cite{Birmingham:1991ty} on the bundle ${\cal A} \longrightarrow {\cal A}/{\cal G}$, and of $\varphi$ as its curvature.

\subsection{Ghost number in combinatorial differential geometry}
\label{CovDer}

Given a differential form on the universal bundle, one can
decompose it  locally in terms of differential forms on $L_{p}$ and
those on ${\cal A}$. The degree of the former is the de Rham
degree, and the degree of the latter the ghost number.

Take the points $x,y,\xi$ that are all infinitesimally close in $L_{p} \times {\cal A}$, such that the projections of $x$ and $y$ onto the second factor are identical, and that the projection of $x$ and $\xi$ onto the first factor are identical. Then the connection $\mu \in \OMEGA^{1}(\mathfrak{P},G)$ can be decomposed as
\be
\mu(x,y) &=& m(x,y) \\
\mu(x,\xi) &=& c(x) ~.
\ee

The BRST transformation $\qq$ is clearly displacement along ${\cal A}$
\be
\tilde\delta_{\mu}^{0} g(x,\xi) &=& \mu(x,\xi)(g(\xi))g(x)^{-1} \\
   &=& c(x)(g(\xi))g(x)^{-1} \\
   &=& c(x)(g(\xi))g(\xi)^{-1} ~g(\xi)g(x)^{-1} ~.
\ee
The last line is really the covariant derivative ``$[c,g] + \qq
g$''. If we drop the Faddeev-Popov ghost setting $c=0$, we get the
covariant exterior derivative on the base space ${\cal P}$ discussed above,
$\qq^{H}$. This means that objects that remain constant in BRST
transformations $\qq$, are covariantly constant sections of ${\cal
A} \longrightarrow {\cal P}$. This will have interesting
repercussions in Sec.~\ref{Abelian}.

\section{The universal gerbe}
\label{UG}

We started in Sec.~\ref{Coho} with a gerbe whose  cohomology class was given by the cocycle data $(\lambda_{ij}, g_{ijk})$ in $\mathbf{G}$. In Sec.~\ref{Fields} we recalled the fields needed to decompose the gerbe fully. Let us denote this set of fields --- the connection data \etc --- for a fixed gerbe in $\mathbf{G}$ by $\hat{{\cal A}}$. This notation is justified, as it is clearly a generalisation of the affine space of connections that appeared in Sec.~\ref{Universal}. 

Let us denote the symmetries of the fully decomposed gerbe in a similar vein  by $\hat{\cal G}$. Then picking a specific fully decomposed gerbe $P_{\mathbf{g}}$ provides a representative for the equivalence class $\mathbf{g} \in {\cal H}=\hat{{\cal A}}/\hat{\cal G}$. The {\em universal gerbe} $\mathfrak{G}$ can be constructed formally (as a set) as the disjoint union of all such representatives of all isomorphism classes of fully decomposed gerbes, and fits in a similar diagram to that of the universal bundle (\ref{UBdiag})
\be
\begin{CD}
P_{\mathbf{g}} @>>> \mathfrak{G} \\
@V{\pi}VV   @VV{\Pi}V \\
X \times \{ g \} @>{i}>> X \times {\cal H} ~.
\end{CD}
\ee 
Again, we keep the cocycle data $(\lambda_{ij}, g_{ijk})$ fixed on a
fixed cover $\{  {\cal U}_{i} \}$ of $X$, and let the connection
data $(m_{i}, \gamma_{ij},B_{i}) \in \hat{\cal A}$ vary freely.
Isomorphism classes in ${\cal H}$ are equivalence classes of
elements of $\hat{\cal A}$ identified by symmetries of a gerbe
$\hat{\cal G}$. 

To show that the quotient $\mathfrak{G} \longrightarrow \mathfrak{G}/\hat{\cal G}$ should actually define a smooth bundle would require careful  topologising $\mathfrak{G}$ and studying the action of $\hat{\cal G}$ on it. As in the case of the universal bundle, the existence of a smooth quotient ${\cal H}=\hat{{\cal A}}/\hat{\cal G}$ would specifically require further assumptions on the gauge data, such as restricting to irreducible connections only. In the discussion that follows, we shall nevertheless need only the fact that ${\cal H}$ provides a local moduli space for fully decomposed gerbes near a fixed reference gerbe, and is not strictly speaking dependent on whether $\mathfrak{G}$ exists as a universal object or not. The local statement is certainly true, though the stronger assertion seems plausible as well. Note that the universal gerbe in the cohomologically Abelian setting of bundle gerbes was defined rigourously in Ref.~\cite{Carey:2001gi}.

We can think of the universal gerbe $\mathfrak{G}$ also as a stack\footnote{Or, more correctly, a stack of categories whose objects are local bundles.} of local universal bundles $\{  {\mathfrak P}_{i} \}$
on $X$, and invertible morphisms between them $\lambda_{ij} \in
\End( {\mathfrak P}_{j} , {\mathfrak P}_{i} )$ with extra structure
$\hat{\cal A}$ and symmetries $\hat{\cal G}$. The symmetries of the gerbe $\hat{\cal G}$ include clearly the structure groups ${\cal G}_{i}$ of the underlying local universal bundles ${\mathfrak P}_{i}$ in a certain way. A mismatch is bound to arise where two universal bundles overlap; the effects of this can be analysed by investigating the behaviour of the horizontal part of the covariant  connection on these bundles in Sec.~\ref{Hori}. 

As in the case of the universal bundle, instead  of the underlying
gerbe $P_{\mathbf{g}} \longrightarrow X \times \{ g\}$ in the equivalence
class $g$, we consider the fully decomposed universal gerbe
$\mathfrak{G} \longrightarrow X \times {\cal H}$ with connection
data $(\mu_{i},V_{ij},A_{i})$. These fields can be expanded in
ghost number
\be
\mu_{i} &=& m_{i} + c_{i} \\
V_{ij} &=& \gamma_{ij} + a_{ij} \\
A_{i} &=& B_{i} + E_{i} + \phi_{i} ~,
\ee
where the lowest components $(m_{i},\gamma_{ij},B_{i})$ are  the
connection data of the underlying gerbe. The higher components
appear in the affine and the gauge transformation data of the
underlying gerbe on $X$; as in the case of the universal bundle,
ghost fields find a natural place in the higher components of the
connection data.  Here only $a_{ij} \in \Omega^{0}({\cal U}_{ij},
\Lie G \otimes \Gset)$ is new in the non-Abelian construction, and
in Sec.~\ref{Abelian} we shall see that it is actually required
for the standard \v{C}ech-de Rham Abelian construction.

In what follows, two bigraded fields with bigrading $(p,q)$
resp.~$(p',q')$ are mutually odd precisely when both the total
gradings $p+q$ and $p'+q'$ are odd. In this way all fields can be
treated consistently as differential forms on the universal gerbe,
rather than differential forms on the underlying manifold with an
additional (ghost number) grading.

The curvatures are defined precisely in the same way as in
Sec.~\ref{Fields},  though now they can be expanded according to
ghost number of each component
\be
F_{i} & = & \nu_{i} + \pi_{i} + \varphi_{i} \\
 &\eq& \F_{i} - \iota_{A_{i}} \\
\Delta_{ij} & = & \delta_{ij} + \eta_{ij} + b_{ij}   \\
&\eq& \lambda_{ij}(A_{j}) - A_{i} + D_{\mu_{i}} V_{ij} - \half [V_{ij},V_{ij}]\\
\Omega_{i}  & = & \omega_{i} + \alpha_{i} + \rho_{i} + \sigma_{i} \\
&\eq& D_{\mu_{i}} A_{i} ~.
\ee
All these fields can be collected in Table \ref{fieldsTable2}.
\begin{table}
\be
\begin{array}{||c|c|cccc||}
    \hline \hline
    & \text{ghost\#}  & \text{0-form} & \text{1-form} &
\text{2-form} & \text{3-form}   \\
    \hline
    G & 0  & g_{ijk} & \gamma_{ij}  & B_i, ~\delta_{ij} & \omega_{i} \\
      & 1  & a_{ij}  & E_i, ~\eta_{ij}&  \alpha_{i}  & \\
      & 2  & \phi_{i}, ~b_{ij} & \rho_i && \\
      & 3  & \sigma_{i} &&& \\
    \hline \Aut(G)
      & 0  & \lambda_{ij} & m_i & \nu_{i}& \\
      & 1  &  c_i    &  \pi_i  && \\
      & 2  &  \varphi_{i}  &&&  \\
  \hline \hline
\end{array}
\nonumber
\ee
\mycaption{Fields and field strengths on the universal gerbe.
\label{fieldsTable2}}
\end{table}

\subsection{The differentials along the universal gerbe}
\label{Longi}

These definitions determine the curvature triple $(\nu_{i},
\delta_{ij}, \omega_{i})$ in terms of the connection data   $(m_{i},
\gamma_{ij}, B_{i})$, as well as the the differentials
\be
\qq m_{i} &=& \pi_{i} + \iota_{E_{i}} - \d_{m_{i}}c_{i} \\
\qq c_{i} &=& \varphi_{i} + \iota_{\phi_{i}}- \half [c_{i},c_{i}] \\
\qq_{c} \gamma_{ij} &=& \eta_{ij}+ E_{i} - \lambda_{ij}(E_{j}) -
\d_{m_{i}}a_{ij} + [\gamma_{ij},a_{ij}] \label{qqgamma} \\
\qq_{c} a_{ij} &=& b_{ij}+ \phi_{i} - \lambda_{ij}(\phi_{j}) +
\half [a_{ij},a_{ij}] \\
\qq_{c} B_{i} &=& \alpha_{i} - \d_{m_{i}}E_{i} \label{qqB}\\
\qq_{c} E_{i} &=& \rho_{i} - \d_{m_{i}} \phi_{i} \\
\qq_{c} \phi_{i} &=& \sigma_{i} ~.
\ee
The form of the Bianchi identities in terms of the universal
connection data  is the same as in  Sec.~\ref{Fields}
\be
D_{\mu_{i}}F_{i} + \iota_{\Omega_{i}} &=& 0  \\
D_{\mu_{i}} \Delta_{ij} +\Omega_{i} - \lambda_{ij}(\Omega_{j}) +
[\iota_{\Delta_{ij}} - F_{i},V_{ij}] &=&  [{\cal C}_{ij}, \lambda_{ij}(A_{j})]  \label{BiDelta} \\
D_{\mu_{i}} \Omega_{i} - [F_{i},A_{i}] &=& 0 ~,
\ee
so that the lowest components in ghost number reproduce precisely
the  corresponding identities on $X$. Note that we keep track of
the constraint functional
\be
{\cal C}_{ij} &\eq& {}^{\lambda_{ij}*}\mu_{j} -\mu_{i} +\iota_{V_{ij}} \\
&=& {\cal C}_{ij}^{0} + {\cal C}_{ij}^{1} ~,
\ee
where
\be
{\cal C}_{ij}^{0} &\eq& {}^{\lambda_{ij}*}
m_{j} -m_{i} +\iota_{\gamma_{ij}} \label{C0} \\
{\cal C}_{ij}^{1} &\eq& {}^{\lambda_{ij}}c_{j} -c_{i} +\iota_{a_{ij}} \label{C1} ~.
\ee
This is because the cocycle equations are needed for an off-shell
construction of the nilpotent derivative $\qq$.
Indeed, the higher components can be used to read off the
differentials
\be
\qq_{c} \pi_{i} &=& -\iota_{\rho_{i}} - \d_{m_{i}}\varphi_{i}  \\
\qq_{c} \varphi_{i} &=& -\iota_{\sigma_{i}} \\
\qq_{c} \eta_{ij} &=& - \d_{m_{i}}b_{ij} - \rho_{i} +
\lambda_{ij}(\rho_{j})+ [\pi_{i} - \iota_{\eta_{ij}}, a_{ij}] +
[\varphi_{i}- \iota_{b_{ij}}, \gamma_{ij}] \nonumber \\
&& - [{\cal C}_{ij}^{0},\lambda_{ij}(\phi_{j})] - [{\cal C}_{ij}^{1},\lambda_{ij}(E_{j})] \\
\qq_{c} b_{ij} &=&  - \sigma_{i} + \lambda_{ij}(\sigma_{j})+
[\varphi_{i} - \iota_{b_{ij}},a_{ij}]  - [{\cal C}_{ij}^{1},\lambda_{ij}(\phi_{j})] \\
\qq_{c} \alpha_{i} &=&  - \d_{m_{i}}\rho_{i} + [\nu_{i},\phi_{i}]
+ [\pi_{i},E_{i}] + [\varphi_{i},B_{i}] \\
\qq_{c} \rho_{i} &=&  - \d_{m_{i}}\sigma_{i} + [\pi_{i},\phi_{i}]
+ [\varphi_{i},E_{i}] \\ \qq_{c} \sigma_{i} &=&
[\varphi_{i},\phi_{i}] ~.
\ee

\begin{theorem}
\label{qqTheorem}
The exterior derivative $\qq$ is an odd,
identically nilpotent (of order two) differential in the field
space.
\end{theorem}
\begin{proof}
Follows immediately from the definition of $\qq$ as the  exterior
derivative on ${\cal H}$ from the point of view of the universal
gerbe $\mathfrak{G} \longrightarrow X \times {\cal H}$. It is
instructive to verify this by a direct calculation as well.
\end{proof}

Modulo a few sign differences, which we shall discuss in detail in
Sec.~\ref{Comparison}, the action of $\qq$ is on-shell the same as
the BRST operator $\Qn$ in Sec.~\ref{Construction} and Theorem
\ref{fullQ} in particular.

\subsubsection{Horizontal derivative}
\label{Hori}

As in Sec.~\ref{Curva}, one may again decompose $\qq = \qqH+\qqV$, where all $c_{i}$ dependence is collected in $\qqV$; this makes $\qqV$ effectively into  a translation along the orbit of local gauge transformations ${\cal G}_{i}$. One may verify that the vertical derivative is still nilpotent $(\qqV)^{2}=0$. The horizontal differential squares to
$(\qqH)^{2} x =[\varphi_{i}+\iota_{\phi_{i}},  ~ x]$ as expected on all other fields than
\be
(\qqH)^{2} \eta_{ij} &=& [\varphi_{i}+\iota_{\phi_{i}}, ~ \eta_{ij} ] + [ {}^{\lambda_{ij}}(\varphi_{j}+\iota_{\phi_{j}}) -(\varphi_{i}+\iota_{\phi_{i}})  , \lambda_{ij}(E_{j})] \label{qqHeta} \\
(\qqH)^{2} b_{ij} &=& [\varphi_{i}+\iota_{\phi_{i}},  ~ b_{ij} ] + [ {}^{\lambda_{ij}}(\varphi_{j}+\iota_{\phi_{j}}) -(\varphi_{i}+\iota_{\phi_{i}})  , \lambda_{ij}(\phi_{j})]   ~.
\ee
The extra piece in (\ref{qqHeta}) is the same that obstructs the nilpotency of $\Qn$ in Theorem \ref{fullQ}.

This calculation shows that we may interpret $\qqH$ as the covariant exterior derivative on $\mathfrak{G} \longrightarrow \mathfrak{G}/{\cal G}_{i}$ only where the local curvature $\varphi_{i}+\iota_{\phi_{i}}$ extends to a well-defined $\Lie \Aut G$-valued section. Outside this domain basic functionals are not necessarily covariantly constant on $\mathfrak{G}/{\cal G}_{i}$. This means effectively that it is not possible to separate the local gauge symmetry part ${\cal G}_{i}$ from the full symmetry group of the gerbe $\hat{\cal G}$ in any clean way when fields on the double intersections ${\cal U}_{ij}$ are taken in general into account. 

In Sec.~\ref{Disc} we shall nevertheless see how the local curvature $\varphi_{i}+\iota_{\phi_{i}}$ does extend to a well-defined $\Lie \Aut G$-valued section at certain physically relevant configurations, namely fixed point loci of the BRST operator, \cf Sec.~\ref{Elim}.

\subsection{Constraint algebra}
\label{Constr}

The BRST transformation rule of $\delta_{ij}$ can be deduced in
two independent ways, on one hand form  the Bianchi identity
(\ref{BiDelta}), on the other by variational calculus from the
definition $\delta_{ij} = \delta_{ij}(m_{i}, \gamma_{ij}, B_{i})$.
The results must be consistent: this leads us to the observation,
as anticipated in Sec.~\ref{Infinites}, that the structure
constants must indeed be held constant in BRST variations $\qq
\lambda_{ij} = 0$, and that the constraint ${\cal C}_{ij}^{1}
\approx 0$ defined in (\ref{C1}) should hold. It follows then $\qq
g_{ijk} = 0$.

The constraint ${\cal C}_{ij} \approx 0$ holds already by
definition of the universal gerbe, where the one-form $\mu_{i}$ is
a part of the connection data and satisfies therefore the
appropriate cocycle conditions. The universal constraints are
indeed defined as follows:
\begin{definition}
\label{UC}
\be
{\cal C}_{ij} &\eq& {}^{\lambda_{ij}*}\mu_{j} -\mu_{i} +\iota_{V_{ij}} \\
{\cal C}_{ijk} &\eq& \partial_{\lambda}V_{ij} +
\delta_{\mu_{i}}^{(0)} g^{-1}_{ijk}  \\
{\cal B}_{ij} &\eq&  {}^{\lambda_{ij}}F_{j}
-F_{i} + \iota_{\Delta_{ij}} \label{dBij} \\
{\cal B}_{ijk} &\eq&  \partial_{\lambda} \Delta_{ij} -
[F_{i},g_{ijk}]   \label{dBijk} ~.
\ee
\end{definition}
The lowest components reproduce
\begin{itemize}
\item     The constraints ${\cal C}_{ij}^{0},{\cal C}_{ijk}^{0},{\cal B}_{ij}^{1}$
 and ${\cal B}_{ijk}^{1}$ of the underlying gerbe as in Sec.~\ref{Triple};
\item     The constraint ${\cal C}_{ij}^{1}$ of (\ref{C1});
\item     The cocycle conditions (\ref{k3}) and(\ref{deltaDiff})
where the former is identically satisfied ${\cal B}_{ij}^{0} = 0$
and the latter is, by Theorem \ref{deltaTheo}, weakly satisfied
${\cal B}_{ijk}^{0} \approx 0$.
\end{itemize}

The new constraints are
\be
{\cal C}_{ijk}^{1} &=& \partial_{\lambda} a_{ij} - [c_{i}, g_{ijk}] \label{nc1} \\
{\cal B}_{ij}^{2} &=&  {}^{\lambda_{ij}}\varphi_{j}
-\varphi_{i} + \iota_{b_{ij}}  \label{nc2} \\
{\cal B}_{ijk}^{2} &=&  \partial_{\lambda} b_{ij} -
[\varphi_{i}, g_{ijk}]  ~. \label{nc3}
\ee
The reason for imposing these constraints is,  again, the geometry
of the universal gerbe. On the other hand, there is
circumstantial evidence already on the level of the underlying
gerbe that they should be imposed: the constraint  ${\cal
B}_{ij}^{2}$ appears as an obstruction in Theorem \ref{fullQ}; the
inner parts of the constraints ${\cal C}_{ijk}^{1}$ and ${\cal
B}_{ijk}^{2}$  follow as   integrality conditions from  ${\cal
C}_{ij}^{1}$ and ${\cal B}_{ij}^{2}$, respectively.

Whether this is an acceptable set of  constraints from the point
of view of the underlying gerbe as well as the universal gerbe
depends on whether it forms, together with the BRST operator
$\qq$, a closed algebra. This can be verified by calculating their
covariant derivatives.
\begin{theorem}
\label{deltaC}
\be
\delta_{\mu_{i}}^{(1)} {\cal C}_{ij} &=&  {\cal B}_{ij} +
[\iota_{V_{ij}} , {\cal C}_{ij}]
\nonumber \\
\delta_{\mu_{i}}^{(1)} {\cal C}_{ijk} &=& {\cal B}_{ijk}  +  [{\cal C}_{ij}, V_{ij}]  + [{}^{\lambda_{ij}}{\cal C}_{jk},  V_{ij} + \lambda_{ij}V_{jk}]   + [{}^{\lambda_{ij}\lambda_{jk}}{\cal C}_{ki},  \partial_{\lambda}V_{ij
} ] \nonumber \\
\delta_{\mu_{i}}^{(2)} {\cal B}_{ij} &=& [\iota_{V_{ij}} , {\cal
B}_{ij}] + [\iota_{A_{i}} + {}^{\lambda_{ij}}F_{j}
 , {\cal C}_{ij} ] \nonumber \\
\delta_{\mu_{i}}^{(2)} {\cal B}_{ijk} &=& [{}^{\iota_{g_{ijk}}}F_{i} , {\cal C}_{ijk}] + [\lambda_{ij} \Delta_{jk} +\lambda_{ij}\lambda_{jk} \Delta_{ki} ,  {\cal C}_{ij}] + [\lambda_{ij}\lambda_{jk} \Delta_{ki} , {}^{\lambda_{ij}}{\cal C}_{jk}] \nonumber \\
&& + [V_{ij} + {\lambda_{ij}}V_{jk}, {\cal B}_{ijk}] -
[{\lambda_{ij}}V_{jk}, {\cal B}_{ij} ] + [\lambda_{ij}\lambda_{jk}
V_{ki} , {}^{\lambda_{ij}\lambda_{jk}} {\cal B}_{ki}] \nonumber
 ~.
\ee
\end{theorem}

From these results it is now possible to read off the actual
constraint algebra. This is because the combinatorial differential
includes the BRST differential $\qq$
\be
\qq_{c} {\cal C}_{ij} &=& \delta_{\mu_{i}}^{(1)} {\cal C}_{ij} - \d_{m_{i}}   {\cal C}_{ij}  - \half [  {\cal C}_{ij},  {\cal C}_{ij}] \\
\qq_{c} {\cal C}_{ijk} &=& \delta_{\mu_{i}}^{(1)} {\cal C}_{ijk} - \d_{m_{i}}   {\cal C}_{ijk} - \half [  {\cal C}_{ijk},  {\cal C}_{ijk}]  \\
\qq_{c} {\cal B}_{ij} &=& \delta_{\mu_{i}}^{(2)} {\cal B}_{ij} - \d_{m_{i}}   {\cal B}_{ij}  \\
\qq_{c} {\cal B}_{ijk} &=& \delta_{\mu_{i}}^{(2)} {\cal B}_{ijk} - \d_{m_{i}}   {\cal B}_{ijk} ~.
\ee
For instance,
\be
\qq_{c} {\cal C}_{ij} &=& {\cal B}_{ij} +
[\iota_{V_{ij}} , {\cal C}_{ij}]
 - \d_{m_{i}}   {\cal C}_{ij}  - \half [  {\cal C}_{ij},  {\cal C}_{ij}] ~.
\ee
This can be decomposed order by order in ghost number
\be
0 &=& {}^{\lambda_{ij}}\nu_{j} - \nu_{i} + \iota_{\delta_{ij}} -
\d_{m_{i}-\gamma_{ij}}   {\cal C}_{ij}^{0}  - \half [  {\cal C}_{ij}^{0},
 {\cal C}_{ij}^{0}]  \\
\qq_{c} {\cal C}_{ij}^{0} &=& {\cal B}_{ij}^{1} +
[\iota_{a_{ij}} , {\cal C}_{ij}^{0}]
 - \d_{m_{i}-\gamma_{ij}}   {\cal C}_{ij}^{1}  -  [  {\cal C}_{ij}^{0},
 {\cal C}_{ij}^{1}] \\
\qq_{c} {\cal C}_{ij}^{1} &=&
{\cal B}_{ij}^{2} +
[\iota_{a_{ij}} , {\cal C}_{ij}^{1}]
 - \half [  {\cal C}_{ij}^{1},  {\cal C}_{ij}^{1}] ~.
\ee
The first of these equations can be checked independently by using
the definitions of $\nu_{i}$ and $\delta_{ij}$ in terms of
connection data. On-shell it reduces to the cocycle condition
(\ref{k3}). The right-hand sides of the rest of the equations
vanish on-shell, and the algebra closes.

There is one final twist to the constraint algebra:  it is still
reducible. This is because one can show again by direct
calculation that the following relationships between the
constraints apply:
\be
\partial_{\lambda}{\cal C}_{ij} &=& \iota_{{\cal C}_{ijk}} \label{r1} \\
\partial_{\lambda}{\cal B}_{ij} &=& \iota_{{\cal B}_{ijk}} \label{r2} ~.
\ee
This means that we must, effectively, include these two equations
in the constraint algebra as further constraints. We do this in
the next section. In that analysis we shall need the following
similar
\begin{lemma}
\label{qqC}
The BRST transformations of the constraints are consistent on triple intersections in the sense 
$ \partial_{\lambda}{\qq\cal C}_{ij} = \iota_{{\qq\cal C}_{ijk}} \label{qC}$.  \qed
\end{lemma}

\subsection{Constraints in the BRST cohomology}
\label{GF}

To trivialise the constraints ${\cal C}_{ij}$, ${\cal C}_{ijk}$,
${\cal B}_{ij}$, and  ${\cal B}_{ijk}$ in BRST cohomology, we need
to introduce two cohomologically trivial  pairs of fields
$(\Lambda_{ij}, K_{ij})$ and $(\Lambda_{ijk}, K_{ijk})$. Expanded
in ghost number, the fields are
\be
\Lambda_{ij} &\eq& \Lambda_{ij}^{-1} + \Lambda_{ij}^{0} \\
K_{ij} &\eq& K_{ij}^{0} +  K_{ij}^{1} ~.
\ee
We can now define their BRST transformations as
\be
\qq  \Lambda_{ij} & \eq & {\cal C}_{ij} -K_{ij} \\
\qq K_{ij} & \eq & \qq_{c}{\cal C}_{ij}
\ee
and similarly for $\Lambda_{ijk},K_{ijk}$. Here $\qq{\cal C}_{ij}$
is a known expression, and reduces on-shell to the constraints
$\qq{\cal C}_{ij} \approx {\cal B}_{ij}^{1} + {\cal B}_{ij}^{2}$.
The  lowest order term $ {\cal B}_{ij}^{0}$ does not, and should
not, appear, as it is algebraically trivial.  The BRST operator
$\qq$ is still nilpotent, and the constraints ${\cal C}_{ij}$ and
${\cal B}_{ij}$ are exact in the cohomology of $\qq$. (Note the
absence of the ghost field $c$ here. Any attempt at making
$(\Lambda_{ij}, K_{ij})$ transform covariantly under $\qq$ would
lead to an accumulation of  $\varphi_{i} + \iota_{\phi_{i}}$ terms
that could not be cancelled.)

The reducibility relations observed in (\ref{r1}) and  (\ref{r2})
can now be taken care of by introducing the ghost-for-ghost fields
$(M_{ijk},N_{ijk})$ and defining
\be
\qq M_{ijk} &=& \partial_{\lambda}\Lambda_{ij} - \iota_{\Lambda_{ijk}} + N_{ijk}\\
\qq N_{ijk} &=& \partial_{\lambda}K_{ij} - \iota_{K_{ijk}} ~.
\ee

We have summarised   the new fields required for trivialising the
constraints in Table \ref{fieldsTable3}.
\begin{table}
\be
\begin{array}{||c|c|ccc||}
    \hline \hline
    & \text{ghost\#}  & \text{0-form} & \text{1-form} &
\text{2-form}    \\
    \hline
    G    & -1  &&\Lambda_{ijk}^{-1}& \\
         &  0  &\Lambda_{ijk}^{0} & K_{ijk}^{0} & \\
         &  1  &K_{ijk}^{1}&& \\
         \hline
\Aut(G)  & -2  && M_{ijk}^{-2} & \\
         & -1  & M_{ijk}^{-1} &\Lambda_{ij}^{-1}, N_{ijk}^{-1} & \\
         &  0  &\Lambda_{ij}^{0}, N_{ijk}^{0} & K_{ij}^{0}& \\
         &  1  & K_{ij}^{1}&& \\
  \hline \hline
\end{array}
\nonumber
\ee
\mycaption{Lagrange multiplies for imposing constraints.
\label{fieldsTable3}}
\end{table}
This table includes fields of so negative ghost number that their
total degree as universal forms is negative, indeed $-1$ for the
components of $M_{ijk}$. From the field theory point of view this
is of no consequence. From the point of view of the universal
gerbe the situation is slightly odd, however, and may suggest that
we should see the \v{C}ech degree as a part of the grading. Then
the total degree of $M_{ijk}$ is zero and $N_{ijk}$ is one.
Similarly the degree of $\Lambda_{ij}, K_{ij}$ is then one and
$\Lambda_{ijk}, K_{ijk}$ is two, and the \v{C}ech differential
$\partial_{\lambda}$ raises  the degree by one.

The constraint algebra closes now, the full BRST operator $\qq$ is
identically  nilpotent, takes into account all the reducibility
relations, and its cohomology is supported on the constraint
surface
\be
{\cal C}_{ij} \approx {\cal C}_{ijk} \approx {\cal B}_{ij} \approx
{\cal B}_{ijk} \approx 0 ~.
\ee
Assuming that we have the traces $\tr_{i}$, $\Tr_{i}$ and the Hodge star $*$ of a Euclidean metric at our disposal (\cf Sec.~\ref{Traces}),  a suitable gauge Fermion that
imposes these constraints in a path integral is
\be
\Psi &=& \Tr_{i} \Lambda_{ij} \wedge * K_{ij}+ \tr_{i}\Lambda_{ijk} \wedge  *
K_{ijk} + \Tr_{i} M_{ijk} \wedge * N_{ijk}   ~.  \label{GFermion}
\ee
Integrating out $N$ one gets a Gaussian suppression for the norm
of $\partial_{\lambda}\Lambda_{ij} - \iota_{\Lambda_{ijk}}$, and
the path integral over $M$ forces $\partial_{\lambda}K_{ij} -
\iota_{K_{ijk}}=0$. $\Lambda$ and $K$ act as Lagrange multipliers
for ${\cal C}$ and ${\cal B}$ respectively.

\section{Comparison}
\label{Comparison}

We have presented in Sec.~\ref{Construction} and \ref{UG} two
very similar constructions that nevertheless differ in certain
detail. To show that they are mathematically equivalent one would
have to demonstrate that the cohomologies of $\Qn$ and $\qq$ are
isomorphic. Of course, as one of the operators, $\Qn$, is {not}
nilpotent this cannot be done directly.

The field space where the nilpotent  operator $\qq$ acts, is
larger than the one where $\Qn$ does. The operators  can,
therefore, be compared only in a locus   where the additional
fields $a_{ij}$ and $b_{ij}$ are somehow eliminated. In a
classical physical theory this could be done by imposing equations
of motion; unfortunately, in want of an action principle, we do
not have enough information to do so, nor should we indeed impose
classical equations of motion on fields which we plan to
quantise.

What we really need to show, in fact,  is that any path integral
with a $\qq$-invariant measure and a $\qq$-invariant integrand,
localises in the new fields $a_{ij}$ and $b_{ij}$ and that the
effective BRST operator $\qq$ acts in this locus as $\Qn$. {\em
This means that the quantum cohomology of $\qq$ is cohomology of
the fully decomposed gerbe} with fixed cocycle data.\footnote{This is cohomology of the fields living on the non-Abelian gerbe, not the cohomology group $H^{1}(X,\mathbf{G})$ of Sec.~\ref{Coho}.} This localisation does indeed happen, and the loci where the path integral localises are the fixed point
loci of $\qq$.

\subsection{Grading}
\label{Grading}

Let us start by eliminating the most obvious difference, namely
that of  grading. In  Sec.~\ref{Construction} the Lie-bracket of
two fields $x$ and $y$ (in a fixed representation) with bigradings $(p,q)$ and $(p',q')$ was
defined
\be
[x,y] &=&
\begin{cases}
xy - (-)^{pp'+qq'} yx & \text{in Sec.~\ref{Construction}} \\
xy - (-)^{(p+q)(p'+q')} yx & \text{in Sec.~\ref{UG}}
\end{cases} ~.
\ee
Also the two BRST operators behaved differently in the presence
of an exterior derivative: for the former we have $\Qn\d = \d\Qn$,  whereas for the latter $\qq\d =- \d\qq$.

We can map the constructions one to the other by
\begin{itemize}
\item[a)]
Mapping every quadratic object $x y$ in the BRST transformation
rules
\be
x y &\mapsto& (-)^{p q'} xy ~,
\ee
where $p$ is the form degree of $x$ and $q'$ ghost number of $y$.
\item[b)]
Redefining fields
\be
(c_i, \varphi_i, \phi_i, \rho_i, \sigma_i) &\mapsto&
(-c_i, -\varphi_i, -\phi_i, -\rho_i, -\sigma_i)  ~.
\ee
\end{itemize}
This mapping is well-defined in the sense that the result does not
depend on the order in which the bilinears are written down. It
also leaves the curvature triple unchanged. There are changes in the
new ghost constraints (\ref{C1}) and (\ref{nc1}) -- (\ref{nc3}).
Applying these rules we get the nilpotent extension $\QQ$ of $\Qn$
\be
\QQ m_{i} &=& \pi_{i} + \iota_{E_{i}} - \d_{m_{i}}c_{i}
\\
\QQ_{c} \gamma_{ij} &=& \eta_{ij}+ E_{i} - \lambda_{ij}(E_{j})
+ \d_{m_{i}}a_{ij} - [\gamma_{ij},a_{ij}]   \\
\QQ_{c} B_{i} &=& \alpha_{i} + \d_{m_{i}}E_{i}   \\
\QQ_{c} \pi_{i} &=& \iota_{\rho_{i}} + \d_{m_{i}}\varphi_{i}
  \\
\QQ_{c} E_{i} &=& -\rho_{i} + \d_{m_{i}} \phi_{i}   \\
\QQ c_{i} &=& \varphi_{i} + \iota_{\phi_{i}} + {\scriptstyle \half} [c_{i},c_{i}]
  \\
\QQ_{c} \eta_{ij} &=& - \d_{m_{i}}b_{ij} + \rho_{i} -
\lambda_{ij}(\rho_{j}) + [\iota_{\eta_{ij}} -\pi_{i} , a_{ij}]
- [\varphi_{i} + \iota_{b_{ij}}, \gamma_{ij}] \nonumber \\
&& + [{\cal C}_{ij}^{0},\lambda_{ij}(\phi_{j})] - [{\cal C}_{ij}^{1},\lambda_{ij}(E_{j})]   \\
\QQ_{c} \alpha_{i} &=&   \d_{m_{i}}\rho_{i} - [\nu_{i},\phi_{i}] - [\pi_{i},E_{i}] - [\varphi_{i},B_{i}]   \\
\QQ_{c} \varphi_{i} &=& -\iota_{\sigma_{i}}   \\
\QQ_{c} \phi_{i} &=& \sigma_{i}   \\
\QQ_{c} \rho_{i} &=&  \d_{m_{i}}\sigma_{i} + [\pi_{i},\phi_{i}] + [\varphi_{i},E_{i}]   \\
\QQ_{c} \sigma_{i} &=& - [\varphi_{i},\phi_{i}]   \\
\QQ_{c} a_{ij} &=& b_{ij}- \phi_{i} + \lambda_{ij}(\phi_{j}) +
{\scriptstyle \half}[a_{ij},a_{ij}]   \\
\QQ_{c} b_{ij} &=&   \sigma_{i} - \lambda_{ij}(\sigma_{j})
-[\varphi_{i} + \iota_{b_{ij}},a_{ij}]  + [{\cal C}_{ij}^{1},\lambda_{ij}(\phi_{j})] \\
\QQ  \Lambda_{ij} & = & {\cal C}_{ij} - K_{ij}   \\
\QQ K_{ij} & = & \QQ_{c}{\cal C}_{ij}   \\
\QQ  \Lambda_{ijk} & = & {\cal C}_{ijk} - K_{ijk}   \\
\QQ K_{ijk} & = & \QQ_{c}{\cal C}_{ijk}   \\
\QQ M_{ijk} &=& \partial_{\lambda}\Lambda_{ij} - \iota_{\Lambda_{ijk}} + N_{ijk}   \\
\QQ N_{ijk} &=& \partial_{\lambda}K_{ij} - \iota_{K_{ijk}}
\ee
This differs from $\Qn$ in the definitions of $\QQ
\gamma_{ij}$  and $\QQ \eta_{ij}$, and in that it involves the
auxiliary fields $a_{ij}$ and $b_{ij}$.

\subsection{On-shell algebra}
\label{OnshellAlg}

The discussion of Sec.~\ref{GF} guarantees that we can make the
path integral localise on subsets of the field space
where the constraints vanish. On that surface we can define an
effective BRST operator $\qqs$ that is formed from $\QQ$ by simply
dropping the constraints that appear explicitly in the definitions
of $\QQ \eta_{ij}$ and $\QQ b_{ij}$
\be
\qqs_{c} \eta_{ij} &=& - \d_{m_{i}}b_{ij} + \rho_{i} -
\lambda_{ij}(\rho_{j}) - [\iota_{\eta_{ij}}+\pi_{i} , a_{ij}] -
[\varphi_{i}+ \iota_{b_{ij}}, \gamma_{ij}] \label{oeta} \\
\qqs_{c} b_{ij} &=&  \sigma_{i} - \lambda_{ij}(\sigma_{j})
-[\varphi_{i} + \iota_{b_{ij}},a_{ij}] ~,  \label{ob}
\ee
and $\qqs x \eq \QQ x$ for any other field $x$. This operator
continues  to be nilpotent  on the constraint surface, as can be
seen using
\begin{lemma}
\label{fullqqs}
\be
\qqs^{2} \gamma_{ij} &=& - [{\cal C}^{0}, \lambda_{ij}(\phi_{j})] +
[{\cal C}^{1}, \lambda_{ij}(E_{j})]  \\
\qqs^{2} \eta_{ij} &=& - [{\cal B}^{1}, \lambda_{ij}(\phi_{j})] +
[{\cal B}^{2}, \lambda_{ij}(E_{j})] + [{\cal C}^{0},
\lambda_{ij}(\sigma_{j})] + [{\cal C}^{1}, \lambda_{ij}(\rho_{j})]  \\
\qqs^{2} a_{ij} &=& -[{\cal C}^{1}, \lambda_{ij}(\phi_{j})] \\
\qqs^{2} b_{ij} &=& - [{\cal B}^{2}, \lambda_{ij}(\phi_{j})] +
[{\cal C}^{1}, \lambda_{ij}(\sigma_{j})] ~,
\ee
and $\qqs^{2}x = 0$ for all other fields.
\end{lemma}
In comparing Theorem \ref{fullQ} and  Lemma $\ref{fullqqs}$ we
notice that the terms involving ${}^{\lambda_{ij}}\varphi_{j}
-\varphi_{i}$ and ${}^{\lambda_{ij}}c_{j} -c_{i}$ in Theorem
\ref{fullQ} have been completed to the constraints ${\cal
B}_{ij}^{2}$ and ${\cal C}_{ij}^{1}$ in Lemma $\ref{fullqqs}$,
respectively. (Other differences have to do with the consistent
treatment and elimination of the new fields $a_{ij}$ and
$b_{ij}$.)

As the original symmetries of the gerbe made use of constraints as
cocycle  conditions, we should  compare $\qqs$ (rather than the
nilpotent $\QQ$) with $\Qn$. What the above discussion shows is
that, on the constraint surface, we can indeed turn $\QQ$
consistently into a non-nilpotent on-shell operator $\qqs$ whose
action generalises, in a certain way, that of $\Qn$.

\subsection{Eliminating auxiliaries}
\label{Elim}

Having dealt with the constraints that appear explicitly in the
definition  of $\QQ$, we are ready to investigate the r\^ole
played by the auxiliary fields $a_{ij}$ and $b_{ij}$. For this we
need the following
\begin{lemma}
\label{FormalFP} Let the odd vector  field $S$ on $V$ be a
symmetry of both the measure $\mu$ and the function $I$. Then the
integral $\int \mu I $ is supported only at the fixed point loci
of $S$ in $V$.
\end{lemma} 
\begin{proof}
The well-known argument \cite{Blau:1995rs} is as follows: let $S=\partial/\partial \theta$ be  an anticommuting vector field on
$V$, and $\theta$ the  local anticommuting coordinate along which
$S$ generates translations. Such a coordinate exists where-ever
the action of $S$ is free.  Let $S$ act freely on $U \subset V$,
so that $\mu =  \mu' \wedge \d \theta $ and $S I =0$. Then
\be
\int_{U} \mu I &=& \int_{U'} \mu' \frac{\partial}{\partial \theta} I = 0
\ee
by the properties of the Berezin integral. Hence the only
nontrivial contributions can arise from the fixed point set  of
$S$ in $V$.
\end{proof}

Requiring that $\qq$ should act consistently on $a_{ij}$, \ie $\qq a_{ij}=0$,  fixes $b_{ij}$ as
a functional of other fields in the theory. At this locus we have
\be
a_{ij} &=& {\veva}_{ij} \label{da} \\
b_{ij} &=&  \phi_{i} - \lambda_{ij}(\phi_{j}) - \half [a_{ij},
a_{ij}]  - [c_{i},a_{ij}] ~, \label{db}
\ee
where ${\veva}_{ij}$ is a fixed  background field $\qq
{\veva}_{ij} =0$. Possible values include, but are not restricted
to, ${\veva}_{ij} =0$. One can check
\be
\QQ \Big( \phi_{i} - \lambda_{ij}(\phi_{j}) - \half [a_{ij},
a_{ij}]  - [c_{i},a_{ij}]\Big) = \QQ b_{ij} ~.
\ee
This means that we can use  (\ref{db}) as an algebraic identity.
By Lemma \ref{fullqqs}, any $\QQ$-invariant path integral then
localises to the values of $a_{ij}$ and  $b_{ij}$ given in
(\ref{da}) and (\ref{db}).

We may now make use of the above values of $a_{ij}$ and $b_{ij}$,
and compare the transformation rules on-shell for $\Qn$ and
$\qqs$. Those that are functionally different are
\be
\qqs \gamma_{ij} &=& \Qn \gamma_{ij} + \d_{m_{i}-\iota_{\gamma_{ij}}}{\veva}_{ij}
  \label{mm1} \\
\qqs \eta_{ij} &=& \Qn \eta_{ij} + [ \iota_{\eta_{ij}}-\pi_{i}  +
\d_{m_{i}-\iota_{\gamma_{ij}}}{\veva}_{ij}, {\veva}_{ij}]
+\d_{m_{i}-\iota_{\gamma_{ij}}}[c_{i},{\veva}_{ij}] \label{mm2}
~.
\ee
When ${\veva}_{ij} =0$ we see that $\qqs$ and $\Qn$ agree.

It is not quite clear from this analysis what r\^ole the other
vacua with ${\veva}_{ij} \neq 0$ play. One possibility is that one
may be able to  localise $a_{ij}$ at $a_{ij}=0$ in the path
integral by suitable gauge fixing. If this is the case, then the
constraint ${\cal C}_{ij}^1$ will force the local Faddeev-Popov
ghosts $c_i$ to form a globally well defined scalar field. This
would mean that   local gauge transformations on different charts
must be globally consistent: the gauge is the same everywhere.

\section{Notes on observables}
\label{Obs}
 
Observables  ${\cal O}$   are BRST-closed $\qq {\cal O}=0$
functionals on the field space. The vacuum expectation values of
BRST-exact functionals vanish. Physical states belong to the
cohomology of $\qq$.  Determining that cohomology is a fundamental problem in Quantum Field Theory.

In this section we look for observables for a fully decomposed non-Abelian gerbe. It turns out that the standard field theory methods do not quite suffice, and the outer part of the automorphism group plays a crucial r\^ole.

\subsection{Local traces}
\label{Traces}

Due to the freedom to choose the frame in the local gauge symmetry, observables ${\cal O}$ should first of all not carry bare indices in representations of $G$ or $\Aut G$. This is because no covariant quantity $x$ is BRST-closed: $\qq x = -[c, x] + \cdots$ does not vanish identically.

Given on each chart ${\cal U}_{i}$ a finite dimensional linear representations of $G$ and $\Aut G$  --- or indeed of the local groups $G_{i}$ and $\Aut G_{i}$, to be more precise ---
we have the traces
\be
\tr_{i} &:& G_{i} \longrightarrow \Rset \\
\Tr_{i} &:& \Aut G_{i} \longrightarrow \Rset
\ee
at our disposal. Traces are not invariant in outer automorphisms, so this does not provide, directly, local invariants for a given field configuration. Since we are at liberty to define each trace locally as we please, we may nevertheless choose them to be compatible in the following sense:
\be
\tr_{i} \lambda_{ij}(x_{j}) &=& \tr_{j} x_{j} ~,
\ee
and similarly for $\Tr_{i}$. It would {\em not} have been possible to assume them to be invariant under arbitrary automorphisms --- the $\lambda_{ij}$ are special.  The cyclic property of the finite dimensional trace guarantees that these choices can be done in a globally consistent way
\be
\tr_{i} \lambda_{ij}  \lambda_{jk}  \lambda_{ki} (x_{i}) =
\tr_{i} \left({g_{ijk}} ~ x_{i} ~g_{ijk}^{-1}\right) = \tr_{i} x_{i} ~,
\ee
and similarly for $\Tr_{i}$. In traditional Quantum Field Theory typical observables are indeed ``invariant\footnote{Invariant under {\em inner} automorphisms.}'' polynomials, \ie combinations of traces of covariant operators, such as Chern classes.

In the pure Yang-Mills case the BRST operator $\qq$ reduces to the covariant exterior  derivative $\qqH$ on the bundle ${\cal A} \longrightarrow {\cal A}/{\cal G}$ whose fibre is the gauge group ${\cal G} = \OMEGA^{0}(X, G)$. The curvature of this differential is one of the scalar fields in the theory, and hence nontrivial. Nevertheless, operated on invariant polynomials on the base space ${\cal A}/{\cal G}$, $\qqH$ is  nilpotent --- thanks to the fact that traces of commutators vanish $\phi \in {\cal G}$
\be
(\qqH)^{2} \tr x &=& \tr [\phi, x] =0 ~.
\ee
In the context of a non-Abelian gerbe, this does not happen, for several reasons:
\begin{itemize}
\item[{\bf (i)}] Invariance does not imply well-definedness on intersections, as even the curvature triple may jump there, \cf (\ref{deltaDiff}), (\ref{k3}), and (\ref{k4}).
\be
{}^{\lambda_{ij}}F_j - F_i & \approx & - \iota_{\Delta_{ij}} \\
\partial_{\lambda_{ij}}(\Delta_{ij}) & \approx & [F_i, g_{ijk}] \\
\lambda_{ij}(\Omega_j) - \Omega_i & \approx & \d_{\mu_i- \iota_{V_{ij}}}(\Delta_{ij})+[V_{ij}, F_i]
\ee
\item[{\bf (ii)}]
The curvature of $\qqH$ is  given locally on ${\cal U}_{i}$ by $\varphi_{i}+\iota_{\phi_{i}}$, but since $\varphi_{i}$ is not  an inner automorphism there is no guarantee that the square $(\qqH)^{2}$ should vanish on traces
\be
(\qqH)^{2} \tr_{i} x_{i}  &=& \tr_{i}[\varphi_{i}+\iota_{\phi_{i}}, x_{i}]  = \tr_{i}[\varphi_{i}, x_{i}] \not\equiv 0~.
\ee
\item[{\bf (iii)}]
Gauge structure is not global; covariant derivatives $\qqH$ on different charts do not glue together consistently on ${\cal U}_{ij}$, \cf Sec.~\ref{Hori}.
\end{itemize}

On the other hand, {\em it is precisely these complications that make it possible for outer automorphisms to appear in BRST cohomology, and to make contact with the cohomology of non-Abelian gerbes in Sec.~\ref{Coho}.} Despite these difficulties, traces have the following two useful  properties:
\begin{lemma}
\label{nice}
\item
Cyclicity of the finite dimensional trace implies
\be
\tr_{i} \d_{m_{i}} \lambda_{ij} X_{j} &\approx& \tr_{i}  \lambda_{ij}
\d_{m_{j}}  X_{j} ~.
\ee
\item
When the connection one-form  is inner, \ie  $m_i = \iota_{n_{i}}$ for some $n_{i} \in \Omega^{1}({\cal U}_{i}, G)$, 
\be
\d  \tr_{i} X_{i} &=&  \tr_{i}  \d_{\iota_{n_{i}}} X_{i} ~. \label{dtr}
\ee
\end{lemma}
\begin{proof}
The first point follows upon using the constraint ${\cal C}_{ij} \approx 0$ and the fact group-valued one-form $\tr_{i}[\gamma_{ij},X_{i}]=0$.
\end{proof}
For general  $\Aut G$-valued forms $m$ (\ref{dtr}) is not true as
$\tr_{i}[m,X]$ does not necessarily vanish. Traces of commutators
vanish only when the automorphism $m$ happens to be inner $m \in
\im \iota$ and its form degree  positive.

\subsection{Fake curvature and Donaldson-Witten invariants}
\label{Encore}

The natural generalisation of the second Chern class that appeared in Donaldson-Witten theory is to replace $\kappa({m}_{i})$ with the fake curvature $\nu_{i}$ and use the local trace
\be
\begin{array}{ll}
{{\half} \Tr_{i} F_{i} \wedge F_{i}  } \\
\quad  = \Tr_{i} \Big( {\half} \nu_{i} \wedge \nu_{i}   +   \pi_{i}  \wedge \nu_{i} +  \varphi_{i}  \wedge \nu_{i} +{\half} \pi_{i}  \wedge \pi_{i} + \pi_{i} \wedge \varphi_{i} + \varphi_{i}  \wedge \varphi_{i} \Big)~.
\end{array}
\label{encoreDW}
\ee
This can be thought of as a local deformation of the Donaldson-Witten invariants by $\iota_{B_{i}}$.
Unlike the Chern class used in Donaldson-Witten theory, (\ref{encoreDW}) does not determine an element in $H^{*}(X,\Rset)$, however:
\begin{itemize}
\item It is not globally defined, unless $\iota_{\Delta_{ij}}$ vanishes. This is because
\be
{}^{\lambda_{ij}}F_{j} &=& F_{i} - \iota_{\Delta_{ij}} ~.
\ee
\item It is not closed, unless $\iota_{\Omega_{i}} $ vanishes
\be
 (\d + \qq) \half ( \Tr_{i} F_{i} \wedge F_{i}  ) &=& -    \Tr_{i} (\iota_{\Omega_{i}} \wedge F_{i} )  ~. \label{tri}
\ee
\end{itemize}
Note that the right-hand side in (\ref{tri}) is a globally on $X$ defined differential form precisely when (\ref{encoreDW}) is. But then (\ref{tri}) is  cohomologically trivial and does not lead to  interesting observables.
The Donaldson-Witten invariants are produced, in fact, only in the essentially Abelian case where
\be
\iota_{\Omega_{i}} = \iota_{\Delta_{ij}} = 0 ~.
\ee
This can be of course arranged by assuming
\be
\iota_{A_{i}} = \delta^{(1)}_{\mu_{i}}(-\iota_{V_{ij}}) = 0 ~.
\ee

\subsection{Abelian cases}
\label{Abelian}

In this section we shall investigate the cohomology of the trace part of a non-Abelian gerbe: this leads to the Abelian gerbe with structure of \cite{Bryl}.\footnote{Note that when the  $\lambda_{ij}$ part of the cocycle data is trivial, the gerbe is called Abelian in \cite{BreenAst}. Indeed, this implies $\iota_{g_{ijk}}=0$. A fully decomposed Abelian gerbe is discussed in detail in \Par 7.3 of Ref.~\cite{BM2}.}

Suppose $m_{i}$ is in the image of $\iota$  so that Lemma
\ref{nice} holds. Let us consider the trace parts of the rest of
the connection data
\be
\bar{B}_{i} &\eq& \tr_{i} B_{i} \\
\bar{A}_{ij} &\eq& \tr_{i} \gamma_{ij} \\
\bar{g}_{ijk} &\eq& {\det}_{i} \, g_{ijk}
\ee
The corresponding three-form  $\bar{\omega}_{i} \eq \d \bar{B} =
\tr_{i}\d_{m_{i}}B_{i}$ is now by (\ref{k2}) closed, and satisfies
by (\ref{k4})
\be
\bar{\omega}_{i}  &=&  \bar{\omega}_{j}+ \d \bar{\delta}_{ij}
\ee
where again $\bar{\delta}_{ij} =  \tr_{i}{\delta}_{ij}$. As long
as there is no more information about $\bar{\delta}_{ij}$, the
local three-forms do not patch together in any useful way.

Suppose further that $\bar{\delta}_{ij} = 0$. Then
$\bar{\omega}_{i}$ extends   to a well-defined global differential
form  $\bar{\omega} \in \Omega^{3}(X,\Rset)$. (Note that this
implies  $\QQ\bar\delta_{ij}=0$, which leads   to further
conditions between ghost fields   $\bar\alpha_{j} -\bar\alpha_{i}
+ \d \bar\eta_{ij} \equiv 0$ modulo traces of commutators.)  The
cocycle conditions (\ref{deltaDeff}), (\ref{gammaDiff}), and
(\ref{funco1}) take  the form
\be
\bar{B}_{j} - \bar{B}_{i} + \d \bar{A}_{ij} &=&  0 \label{a3} \\
\bar{A}_{ij} + \bar{A}_{jk} + \bar{A}_{ki} - \d \ln \bar{g}_{ijk}
&=& 0 \label{a2} \\
\bar{g}_{jkl} \bar{g}_{ijl} \bar{g}_{ijk}^{-1} \bar{g}_{ikl}^{-1}
&=& 1 \label{a1} ~,
\ee
where we used $\tr_{i}\delta^{(0)}g_{ijk}^{-1} =  - \d \ln
\bar{g}_{ijk}$. This defines a representative of a class
$[\bar{B}_{i}, \bar{A}_{ij}, \bar{g}_{ijk}^{-1}]$ in the standard
\v{C}ech-de Rham cohomology or, in other words, an Abelian gerbe
with connection and curving \cite{Bryl}.

It is interesting to find the part of the symmetries of the
non-Abelian gerbe that correspond to the standard action of a
\v{C}ech-de Rham one-cochain on the above two-cocycle.

The symmetries of the non-Abelian gerbe involve among other
generators $\bar{E}_{i} \eq \tr_{i} E_{i}$ and $\bar{a}_{ij} \eq
\tr_{i} a_{ij}$. As the one-cochain involves real fields and not
ghosts, we need to consider $E_i, a_{ij}$ as elements of
$\Omega^1({\cal U}_{i},\Lie G)$ and $\Omega^0({\cal U}_{ij},\Lie G)$. As we have argued
in Sec.~\ref{Grading}, this change of grading forces us to change
the sign in front of all exterior derivatives $\d
\longrightarrow -\d $. With this understanding,
(\ref{qqgamma}) and (\ref{qqB}) lead to
\be
\bar{B}_{i}'  &=& \bar{B}_{i} + \d \bar{E}_{i} \label{s1} \\
\bar{A}_{ij}' &=&  \bar{A}_{ij} - \bar{E}_{j} + \bar{E}_{i} +
\d  \bar{a}_{ij}  \label{s2} \\
\ln \bar{g}_{jkl}'   &=& \ln \bar{g}_{ijk}+\bar{a}_{ij} +
\bar{a}_{jk} + \bar{a}_{ki}\label{s3}  ~.
\ee
The two first rules can be read off, of course, directly from the definition of $\QQ$ as well.

The last transformation rule (\ref{s3}) may appear surprising,
however, given that $\QQ$  annihilates all cocycle data
$(\lambda_{ij}, g_{ijk})$; yet it is required to keep (\ref{a2})
invariant. It can be derived as follows: if we want to compare the
values of a group-valued function $g(x)$ at two different points
on $x,y \in X$ we have to parallel transport the group element
from one point to the other {\em covariantly} to be able to
perform the comparison. The difference is then given precisely by
the combinatorial derivative $m(x,y)(g(y)) g(x)^{-1} =
\tilde\delta^{(0)}_m g(x,y)$.

The same is true of comparing the values of the group  element in
different points $x,\xi$ on the universal gerbe $\mathfrak{G}$ on
the same orbit of the action of the symmetry group $\xi \in
\hat{\cal G} \cdot x$. As the group element $g_{ijk}$ is constant
$\QQ g_{ijk} (x,\xi) = \tilde\delta^{(0)} g(x,\xi) = 1$ in these
transformations, we have $g_{ijk}(\xi) = g_{ijk}(x)$.
Nevertheless, the frame in $\mathfrak{G}$ changes along the way
due to the presence of the curvature of the connection
$\mu_{i}(x,\xi) = -c_i(x)$ so that
\be
g_{ijk}' g_{ijk}^{-1} &=& \mu_{i}(x,\xi)(g_{ijk}(\xi)) g^{-1}_{ijk}(x)\\
&=&  \tilde\delta^{(0)}_\mu(g_{ijk})(x,\xi) \\
&=& -c_{i}(g_{ijk}) g_{ijk}^{-1} ~.
\ee
The part of the symmetry group that is responsible  for this
change is clearly the group of local gauge transformations ${\cal
G}_{i}\subset \hat{\cal G}$. As discussed in Sec.~\ref{Curva}, the
(locally defined) covariant exterior derivative $\qqbarH$ on the base space $\mathfrak{G}/{\cal G}_{i}$ can be obtained from $\QQ$ formally by setting $c_{i}=0$. (Note that the fields $\eta_{ij}$ and $b_{ij}$ that caused trouble in Sec.~\ref{Hori} should here be set to trivial values.) This means that on this base space $g_{ijk}$ is covariantly constant $\qqbarH g_{ijk}=0$. The extra terms in (\ref{s3}) appear therefore as a consequence of eliminating the non-Abelian symmetry ${\cal G}_{i} \subset \hat{\cal G}$, and restricting to basic cohomology 
on $\mathfrak{G}/{\cal G}_{i}$. 

The Abelian part transforms then
\be
\ln \bar{g}'_{ijk} &=& \ln \det -c_i(g_{ijk}) \\
&=& \ln \bar{g}_{ijk}  + \ln \det [-c_i, g_{ijk}] \\
&=& \ln \bar{g}_{ijk}  + \tr_i [-c_i, g_{ijk}] \label{step1} \\
&\approx& \ln \bar{g}'_{ijk}  + \partial_\lambda \bar{a}_{ij}
\label{step2} ~.
\ee
We have used  at (\ref{step1}) the fact that $c_i$ is really a
one-form, and  at (\ref{step2}) the constraint ${\cal C}_{ijk}^1
\approx 0$. Similarly, under $\bar{\phi}_{i} \eq \tr_{i} \phi_{i}$,
\be
\bar{E}_{i}' &=& \bar{E}_{i} +\d \bar{\phi}_{i} \\
\bar{a}_{ij}' &=& \bar{a}_{ij} + \bar{\phi}_{j}  - \bar{\phi}_{i}  ~.
\ee

There are two obvious candidates for observables, but both fail to
be BRST-closed unless we impose conditions on $\bar\alpha_{i}$ and
$\bar\eta_{ij}$.
\begin{itemize}
\item
The three-form $\omega_i$. It fails to be closed by $\QQ \bar{\omega}_{i}
= \d \bar{\alpha}_{i}$.
\item
Given a triangulation with sides $s$, edges $e$ and vertices $v$
of a three-dimensional surface $M$, we can define the holonomy \cite{Bryl}
\be
\hol_{[\bar{B}_{i}, \bar{A}_{ij}, \bar{g}_{ijk}^{-1}]} &=&
\sum_{s \subset M} \int_{s} \bar{B}_{s} + \sum_{e \subset \partial s}
 \int_{e} \bar{A}_{es} +  \sum_{v \in \partial e}  \ln \bar{g}_{ves}^{-1} \label{holo}
\ee
which transforms by the holonomy
\be
\hol_{[\bar{E}_{i}, \bar{a}_{ij}]} &=&  \sum_{e \subset \partial M}
\int_{e} \bar{E}_{e} -  \sum_{v \in \partial e}  \bar{a}_{ve}
\ee
under Abelian symmetries, but picks up the extra piece
$\hol_{[\bar{\alpha}_{i}, \bar{\eta}_{ij},1]}$ under non-Abelian
symmetries.  The former transformation vanishes on closed surfaces
$\partial M = 0$ whereas the latter does not.
\end{itemize}

\section{Discussion}
\label{Disc}

We have proposed two equivalent constructions for a nilpotent  BRST operator $\QQ$ and $\qq$ that both generate  infinitesimal symmetries of a non-Abelian gerbe, though on differently-graded differential forms. For this it was crucial to arrange the cocycle conditions of \cite{BM2} in two categories: 
\begin{itemize}
\item The constraints that the gauge potentials in connection data satisfy;   
\item The Bianchi identities that the curvature triple satisfy on-shell. 
\end{itemize} 
This was possible, as the curvature triple turned out to be completely determined once the connection data was given. 

This is exactly what is needed for defining a path integral measure in quantising the theory as well: the measure can now be easily written down by integrating over all free fields (connection data, affine data, Lagrange multipliers) and imposing the constraints with the help of a gauge Fermion, such as (\ref{GFermion}). Having thus defined the measure, we are nevertheless still lacking a local invariant action principle that would lead to a finite path integral, and well-defined correlators for observables. 

It would now be interesting to determine the BRST cohomology in terms of functionals composed of fields living on the gerbe. Standard methods in QFT do not seem to be able to catch the special features associated with the crossed module $G \longrightarrow \Aut G$ but tend to collapse it to an Abelian $\Z G$ gerbe. There are indeed three crucial differences to traditional Topological Quantum Field Theory
\begin{itemize}
\item Traces of commutators such as $\tr_{i}[m_{i},B_{i}]$ do not vanish, unless both operators are group-valued;
\item Traces of differential forms are invariant polynomials only in the sub-sector of the theory where $\varphi_{i}$ is in the image of $\iota$, \ie it is an inner automorphism;
\item Locally invariant polynomials are not necessarily globally invariant, if they involve either $\eta_{ij}$ or $b_{ij}$.
\end{itemize}
The BRST algebra we have found is not affected directly by any of these phenomena. However, it is precisely these features that are sensitive to the effects of the outer part of the automorphism group $\Out G$, and are likely to make it possible to recover some of the structure of the underlying cohomology of the gerbe $H^{1}(\mathbf{G})$.

In standard Yang-Mills theory, gauge invariant observables were easily identified as elements of the basic complex, and the BRST operator turned out to be the associated covariant derivative. This structure is repeated here only outside double intersections. On double intersections the action of the horizontal BRST operator \eg  on $\eta_{ij}$ contains extra pieces that do not have the interpretation as a curvature. If one nevertheless restricts to configurations where the na\"ive curvature is fully covariant ${}^{\lambda_{ij}}(\varphi_{j}+\iota_{\phi_{j}}) = \varphi_{i}+\iota_{\phi_{i}}$, the mismatch vanishes. 

Neither $c_{i}$, nor $\varphi_{i}$, nor ${\phi_{i}}$ can in general be assumed to extend to an everywhere well-defined object. To keep track of these mismatches in local gauge structure, we had to introduce the new fields $a_{ij}$ and $b_{ij}$ that were not present in the original fully decomposed gerbe. At the fixed point locus of the BRST operator it turned out that $b_{ij}$ was essentially the failure of $\phi_{i}$ to extend to a global section, and that the constraint ${\cal B}_{ij}^{2} \approx 0$ then effectively guaranteed --- again, only at the fixed point locus $a_{ij}=0$ --- that $\varphi_{i}+\iota_{\phi_{i}}$ should indeed transform covariantly from one chart to an other with $\lambda_{ij}$.  At this locus we can define basic functionals that are invariant under local gauge transformations (though only under inner automorphisms), and quotient out consistently the inner part of the local gauge groups ${\cal G}_{i}$. 

The mismatch in $c_{i}$ was measured in terms of $\iota_{a_{ij}}$. This field was required in Sec.~\ref{Abelian} for realising the Abelian gerbe's symmetries consistently. At the fixed point locus we could choose any fixed background value $a_{ij} = \tilde{a}_{ij}$, though the trivial value $a_{ij}=0$ was the one that reproduced the BRST operator of the non-Abelian gerbe. It remains an interesting problem to understand the significance of these other fixed point loci.

The use of combinatorial differential geometry simplified further certain standard operations in BRST quantisation. For instance, ghost number grading is easy to implement in terms of combinatorial differential geometry, this lead to insights in the gauge structure that would otherwise be rather difficult to achieve. This became particularly obvious in the calculation of the constraint algebra, and in extracting the Abelian \v{C}ech-de Rham structure. 

The present structure differs in fact from direct generalisations of 
the \v{C}ech-de Rham treatment of Abelian gerbes such as \cite{Kalkkinen:1999pm} for instance through  the presence of $\delta_{ij}$. Only setting this part of curvature to zero do we get the familiar relationship between a jump in the $B_{i}$-field and the exterior derivative of a one-form $\gamma_{ij}$. 
Furthermore, in the present considerations the analogue of 
the \v{C}ech coboundary operator $\partial_{\lambda}$ did not change the grading or the degree of the fields on which it operated. This is in contrast with the Abelian case, where the connection and the curving of a gerbe fit in a  \v{C}ech-de Rham cocycle where the \v{C}ech and the de Rham form degree are on equal footing. It was only in discussing the ghost number assignments of the Lagrange multipliers for the constraints that it seemed reasonable to take \v{C}ech degree to contribute to the total grading. 

\bigskip 

Finally, it would be interesting to calculate the cohomology of the BRST operator and to compare it to the cohomology of the underlying gerbe. Also, a non-trivial action principle for path integral quantisation is still lacking. The results presented here will hopefully open doors for making use of these structures more directly in String and Quantum Field Theory, \cf \cite{Kalkkinen:2005xg}. Possible applications where the r\^ole of the automorphism group comes to its full right are situations where local perturbative descriptions of a quantum field theory differ globally by non-perturbative symmetry operations, \eg in non-geometric backgrounds of String Theory.

\subsubsection*{Acknowledgements} 

I am indebted to Larry Breen for many enjoyable discussions and  patient guidance. I would also like to thank Bernard Julia and Christiaan Hofman for discussions. This research is supported by a Particle Physics and Astronomy Research Council (PPARC) Postdoctoral Fellowship.

\end{document}